\PassOptionsToPackage{dvipsnames}{xcolor}

\documentclass[pmlr,twocolumn,10pt]{jmlr} 

\usepackage{longtable}

\usepackage{booktabs}
\usepackage{csquotes}
\usepackage{float}
\usepackage{wrapfig}
\usepackage{placeins}

\usepackage[load-configurations=version-1]{siunitx} 

\makeatletter
\def\set@curr@file#1{\def\@curr@file{#1}} 
\makeatother

\theorembodyfont{\upshape}
\theoremheaderfont{\scshape}
\theorempostheader{:}
\theoremsep{\newline}

\jmlryear{2024}
\jmlrworkshop{Conference on Health, Inference, and Learning (CHIL) 2024} 

\DeclareUnicodeCharacter{03B3}{$\gamma$}

\newcommand\ie{i.\,e.\xspace}
\newcommand\eg{e.\,g.\xspace}

\newcommand{\model}{MHMMX\xspace}
\renewcommand{\models}{MHMMXs\xspace}

\usepackage{pifont}
\newcommand{\cmark}{\textcolor{ForestGreen}{\ding{51}}}
\newcommand{\xmark}{\textcolor{BrickRed}{\ding{55}}}

  \DeclareMathOperator*{\argmax}{arg\,max}
  \newcommand{\abs}[1]{\left\lvert #1 \right\rvert}
  \newcommand{\norm}[1]{\left\lVert#1\right\rVert}

\usepackage{tikz}
  \usetikzlibrary{bayesnet}
  \usetikzlibrary{arrows}

\title[Data-driven subgrouping of chronic diseases]{Data-driven subgrouping of patient trajectories with chronic diseases: Evidence from low back pain}

\author{
\Name{Christof Naumzik} \Email{christof.naumzik@googlemail.com}\\
\addr ETH Zurich, Switzerland
\AND
\Name{Alice Kongsted} \Email{akongsted@health.sdu.dk}\\
\addr University of Southern Denmark \& Chiropractic Knowledge Hub, Denmark
\AND
\Name{Werner Vach} \Email{werner.vach@basel-academy.ch}\\
\addr Basel Academy for Quality and Research in Medicine \& University of Basel, Switzerland
\AND
\Name{Stefan Feuerriegel} \Email{feuerriegel@lmu.de}\\
\addr LMU Munich \& Munich Center for Machine Learning, Germany
}

\allowdisplaybreaks

\begin{document}

\maketitle

\begin{abstract}
Clinical data informs the personalization of health care with a potential for more effective disease management. In practice, this is achieved by \emph{subgrouping}, whereby clusters with similar patient characteristics are identified and then receive customized treatment plans with the goal of targeting subgroup-specific disease dynamics. In this paper, we propose a novel mixture hidden Markov model for subgrouping patient trajectories from \emph{chronic diseases}. Our model is probabilistic and carefully designed to capture different trajectory phases of chronic diseases (i.e., ``severe'', ``moderate'', and ``mild'') through tailored latent states. We demonstrate our subgrouping framework based on a longitudinal study across 847 patients with non-specific low back pain. Here, our subgrouping framework identifies 8 subgroups. Further, we show that our subgrouping framework outperforms common baselines in terms of cluster validity indices. Finally, we discuss the applicability of the model to other chronic and long-lasting diseases. 
\end{abstract}

\paragraph*{Data and Code Availability}

Data and code are publicly available via \url{https://github.com/sfeuerriegel/PatientSubgrouping}. To date, our data is one of the largest public datasets with longitudinal information about chronic disease progression.  

\paragraph*{Institutional Review Board (IRB)}

The data used in this study was described previously \citep{Kongsted.2015}. The data collection followed the Declaration of Helsinki, the guidelines of good clinical practice, local laws, and the ordinance on clinical research. Due to absence of invasive tests and interventions, no ethics approval was required by Danish law. Nevertheless, the Regional Ethics Committee for Southern Denmark was consulted about the study \citep{Kongsted.2015}. 

\section{Introduction}

Personalized care aims at better matching patients to treatment plans. In clinical practice, this is commonly achieved through subgroup-specific care \citep{Ashley.2015,Hamburg.2010,Hill.2020,Horwitz.2018,Schleidgen.2013}.

\begin{table*}[h]
\centering
\tiny
\begin{tabular}{ll llllll p{4.5cm}}
\toprule
Method & Approach & Latent &  Probab. & Static & Time & Assignm. & Inter- & Examples \\ 
 &  & states & & var. & series &  & pret. &  \\ 
\midrule
$k$-means & Distance & \xmark & \xmark & \cmark & \xmark & Offline & \cmark & Heart failure \citep{Panahiazar.2015} \\
HC & Distance & \xmark & \xmark & \cmark & \xmark & Offline &\cmark & Metabolism states  \citep{Cohen.2010} \\ 
HMM+PAM & Distance & \xmark &  \xmark & \xmark & \cmark & Online & \xmark & Health self-reports \citep{Ghassempour.2014} \\ 
SOM-VAE & Distance & \xmark &  \xmark & \xmark & \cmark & Online & \xmark & Intensive care units \citep{Fortuin.2019} \\
T-DPSOM & Distance & \xmark &  \cmark & \xmark & \cmark & Online & \xmark$^\dagger$ & Intensive care units \citep{Manduchi.2021} \\
\midrule
LCA & Model & \cmark &  \cmark & \cmark & \xmark & Offline & \cmark & Low back pain \citep{Downie.2016,Kongsted.2015,Nielsen.2016} \\ 
LCA (two-staged)  & Model & \cmark &  \cmark & \cmark & \xmark & Offline & \cmark & Low back pain \citep{Nielsen.2016} \\ 

\midrule
\model (\textbf{ours}) & Model & \cmark & \cmark & \cmark & \cmark & Both & \cmark & Low back pain  \\ 
\bottomrule
\multicolumn{9}{p{15cm}}{\tiny $^\dagger$ We adopt the taxonomy of explainability vs. interpretability from \citet{Rudin.2019}. The model allows for post-hoc explainability but not interpretability of the coefficients and thus of the decision logic (due to the deep VAE).} \\
\multicolumn{9}{p{15cm}}{\tiny HC: hierarchical clustering; HMM: hidden Markov model; LCA: latent class analysis; PAM: partitioning around medoids (PAM); SOM: self-organizing map; VAE: variational autoencoder}
\end{tabular}
\vspace{-0.3cm}
\caption{Examples of patient subgrouping methods in medicine.} 
\label{tbl:overview_subgrouping}
\vspace{-0.5cm}
\end{table*}

The objective in \textbf{\emph{subgrouping}} is to identify clinically-relevant cohorts of patients with similar disease progression. The underlying approach must closely align with clinical practice: modern electronic health data not only store risk factors (\eg, sociodemographics, symptoms, general health profile) but also provides an opportunity to collect longitudinal health trajectories that capture the actual disease progression over time. However, existing approaches for subgrouping primarily make only use of {either} risk factors {or} health trajectories \citep[\eg,][]{Cohen.2010,Downie.2016,Fortuin.2019,Ghassempour.2014,Kongsted.2015,Manduchi.2021,Nielsen.2016,Panahiazar.2015}. Motivated by this, we develop a rigorous, clinically-informed, and probabilistic approach for subgrouping.

In this work, we develop a {probabilistic}, data-driven approach to subgrouping health trajectories that is tailored to \emph{chronic diseases}. Specifically, we account for the different trajectory phases of chronic diseases (i.e., ``severe'', ``moderate'', and ``mild'') through a tailored latent state model. To this end, we develop a novel mixture hidden Markov model~(\model) for clustering patients based on their risk factors \emph{and} their health trajectories. 

Our model fulfills several desirable characteristics for clinical practice aimed at monitoring and treating \emph{chronic diseases}: (1)~Our model is informed by domain knowledge in medicine \cite{Corbin.1988,Corbin.1991}, whereby chronic diseases undergo different trajectory phases. We account for the trajectory phases through a latent state. (2)~Our model supports both static and dynamic variables, which correspond to risk factors at baseline and health trajectories over time, respectively. (3)~Our model is interpretable (e.g., one can interpret the coefficients in the subgroup assignment). This is important to be able to identify clinically-relevant characteristics that are unique to a specific subgroup and thus to generate new disease markers. Such disease markers may later help clinical decision-makers to recognize specific subforms of diseases \citep{strimbu2010biomarkers}. (4)~Our model is probabilistic. This is often beneficial in practice for uncertainty quantification and thus decision-making \citep{Manduchi.2021}. (5)~Our model allows for online/offline assignments. This is helpful in practice where there is not only the goal for subgroup identification but also to assign patients -- both incoming patients and patients with existing health trajectory data -- to their corresponding subgroup.  

For our empirical analysis, we formed an interdisciplinary team with health researchers who aim to provide more effective care to patients with non-specific low back pain. Specifically, we evaluate our model using an extensive longitudinal study of 847 patients with non-specific low back pain. Non-specific low back pain is typically a long-lasting recurrent or persistent condition, and, moreover, is responsible for the most years lived with disability worldwide \citep{GBD.2019}. Patients with low back pain are characterized by considerable heterogeneity, and, still, the between-patient heterogeneity is poorly understood \citep{Silva.2022}, which makes low back pain a compelling use case of direct clinical relevance for demonstrating our model. In particular, our model generates new insights that may prove meaningful for disease management and that may help in identifying new disease markers. 

$\bullet$\,\textbf{Method contributions:} We develop a novel mixture hidden Markov model (\model) for patient subgrouping tailored to chronic diseases. Specifically, we propose a mixture hidden Markov model with a copula approach to model multivariate observations over time. Our model has four key strengths: (i)~It uses both static information (risk factors at baseline) and health trajectories for subgrouping. (ii)~It models the dependence structure in multivariate health trajectories. Note that dependence structure means that different symptoms are typically not independent but that symptoms tend to be jointly present or jointly absent. (iii)~It has a parsimonious and thus interpretable structure. (iv)~It is probabilistic, so that the reliability in assigning patients to subgroups can be examined.

$\bullet$\,\textbf{Dataset contribution:} Further, we contribute a novel, large-scale dataset with patient trajectories from low back pain.

\section{Background}
\label{sec:background}

\subsection{Subgrouping aims}

A common approach to personalized care is \emph{subgrouping} \citep{Ashley.2015,Hamburg.2010,Hill.2020,Horwitz.2018,Schleidgen.2013}. In subgrouping, the aim is to identify clusters of patients with similar disease progression and, then, assign a customized treatment plan for each cluster. In particular, subgrouping is used in clinical practice as a first step to identify new subforms of diseases \cite{Hill.2008} that may have different progression dynamics so that custom treatment plans for this subgroup can be developed. Existing approaches to subgrouping vary along three dimensions (see Table~\ref{tbl:overview_subgrouping}): (1)~data (\ie, what to cluster), (2)~methods (\ie, how to cluster), and (3)~scope (\ie, which type of disease is modeled). These are discussed in the following.

\subsection{Subgrouping approaches}

\textbf{Data:} (i)~\emph{Static information.} There are individual factors determining one's risk and are typically captured in electronic health records as static variables. Examples are sociodemographic variables (\eg, age, gender, presence of health insurance), one's individual health history (\eg, previous conditions and symptoms at presentation), and general health profile (\eg, weight, genetics, smoking). These risk factors are indicative of the future course of a condition and thus help explain the heterogeneity across patient outcomes. Some works even draw upon data from new medical technologies such as MRI scans \citep{Jin.2021}.\footnote{In principle, static variables could be concatenated to dynamic variables. However, this would lead to clustering by patient characteristics (e.g., gender) and \emph{not} by disease dynamics (e.g., relapsing or fluctuating progression), which is not intended in subgrouping for identifying disease markers.} (ii)~\emph{Health trajectories.} One can consider additional factors affecting and reflecting the development of the disease. These are typically given by longitudinal information that captures the health trajectory. Ideally, subgrouping should accommodate both types of data in order to personalize care to patients. However, previous works for subgrouping oftentimes make use of either risk factors \citep[\eg,][]{Cohen.2010,Hill.2008,Nielsen.2016,Panahiazar.2015} or time series data \citep[\eg,][]{Downie.2016,Fortuin.2019,Ghassempour.2014,Kongsted.2015,Manduchi.2021} but rarely both.

\textbf{Methods:} Methods for clustering can be classified into (i)~\emph{distance-based} and (ii)~\emph{model-based} approaches. 
There are also works dealing with \emph{predictive} clustering \citep[\eg,][]{lee2020temporal}, yet which is a different task in which clusters should be determined based on their predictive power of the future outcomes of patients.

\emph{(i)~Distance-based approaches} require that a suitable form of similarity between patients has been defined in order to apply standard clustering algorithms \citep{Cohen.2010,Panahiazar.2015,zhang2023pipeline}. A prominent example of the latter is, \eg, $k$-means clustering. 

\emph{(ii)~Model-based approaches} first fit a model to the input data, which is likely to fit the underlying data-generating process, and then identify clusters based on the estimated model parameters.  \citet{Aghabozorgi.2015} provide an overview of methods for model-based clustering but outside of medicine. For the purpose of subgrouping, a common approach is latent class analysis (LCA), \ie, a finite mixture model, to identify subgroups \citep{Downie.2016,Kongsted.2015,Nielsen.2016}. \citet{Nielsen.2016} have further extended the conventional LCA into a two-staged variant where, in the first stage, risk factors are first grouped into meaningful semantic categories to compute a summary score for each category, and where, in the second stage, the summary scores are clustered using LCA. Here, the premise is that this two-staged approach may allow health professionals to encode domain knowledge.

Related is also group-based trajectory modeling \citep{Nagin2018,Murray2020,Murray2022,Nagin2024}, yet which is typically concerned with single-dimensional observations (not multi-dimensional observations as in our case) and which does not build upon latent state models. However, as we detail below, latent state models are crucial for modeling disease dynamics of chronic conditions.

\textbf{Probabilistic modeling:} In subgrouping, there are important benefits for probabilistic modeling. (1)~Probabilistic models can learn the optimal number of clusters through an information criterion (and not through heuristics such as the elbow curve as in $k$-means). (2)~Probabilistic models avoid simply matching a patient to a single `best' subgroup but assign a subgroup probability, so that practitioners can assess the confidence with which a patient is matched to different subgroups. This allows one to assess the quality of the matching (see Fig.~\ref{fig:cluster_weights}, Appendix~\ref{appendix:online_offline_assignment}).

\vspace{-0.3cm}

\subsection{The HMM-based framework} 
\label{sec:RW_HMM}

Hidden Markov models (HMMs) are a flexible class of time-series models with latent states, in which a time series undergoes transitions between a discrete set of unobservable states \citep{MacDonald.1997,Rabiner.1989}. As such, observations from the time series are noisy realizations of the unobservable (\ie, latent) states. HMMs have become widespread in modeling health trajectories \citep[\eg,][]{DeSantis.2011,gonccalves2023inferring,liu2023identifiability,Maag.2021,EJOR,Scott.2005,Shirley.2010}. One reason is that many time series are driven by latent dynamics, and, hence, HMMs achieve a good fit. Another reason is of theoretical nature. The Corbin-Strauss trajectory framework \citep{Corbin.1988,Corbin.1991,Corbin.1998}, which has found widespread adoption in medical practice and research \citep{Henly.2017,Larsen.2017,RoyalMarsden.2015}, stipulates that chronic, long-lasting, and several other conditions undergo different phases. 

\textbf{Research gap:} A clinically-informed approach to subgrouping that is probabilistic and that combines static and longitudinal information under a dependence structure is missing. Here, we develop a custom mixture hidden Markov model (MHMM) for this purpose. 

\vspace{-0.3cm}
\section{Model development} 
\label{sec:method}

\subsection{Problem statement}

Our objective is to identify subgroups of patients who exhibit similar disease dynamics. Here, we use two sets of variables, both of which are widely available in modern electronic health data: (1)~Static information typically captures risk factors at the initial consultation. These refer to sociodemographics, symptoms, general health profile, etc., denoted by $x_i$ for patient $i$. (2)~Longitudinal information in the form of the health trajectory. Health trajectories refer to how symptoms change over time, denoted by $y_{it}$ for patient $i$ and time step $t = 1, \ldots, T$. 

Our data-driven model for subgrouping is aimed at chronic diseases, and it thus closely adheres to existing medical knowledge on the progression dynamics of \emph{chronic diseases}. To this end, we model the disease progression through the use of latent variables and thus use the hidden Markov model (HMM)-based framework for three reasons. (1)~From a theoretical point, it is widely known that chronic conditions follow the so-called trajectory framework according to which conditions recurrently undergo different states \citep{Corbin.1988,Corbin.1991} with acute and non-acute phases. For example, in low back pain, one would expect different phases characterized as ``severe'', ``moderate'', and ``mild''. Hence, to capture the different phases, we formalize the progression of health trajectories through latent states using HMMs. (2)~From an empirical point, the use of latent states through HMMs has been shown to lead to a better fit \citep{EJOR,AttDMM}. The mathematical reason is that latent states directly assume that symptoms are only noisy realizations. (3)~From a practical point, the latent states in HMMs generally allow for intuitive interpretations \citep{Allam.2021}, and relate to clinical guidelines in practice.

\vspace{-0.5cm}
\subsection{Mixture hidden Markov model for patient subgrouping (\model)}

\subsubsection{Mixture of subgroup HMMs}

The input to the model is as follows. Let $x_i$ denote the risk factors and $y_{it}$ refer to the multivariate time series with symptoms (\eg, consisting of pain $y_{it}^P$ and disability levels $y_{it}^D$) for patient $i$ and time steps $t = 1, \ldots, T$. The values of $y_{it}$ are also called observations or emissions. Let $K$ be the number of unknown clinical subgroups within a specific disease. We refer to the individual subgroups via $k = 1, \ldots, K$. Our objective is to assign patients to these subgroups. Hence, the main output is a probability $\omega_i^k$ with which a patient $i$ belongs to subgroup $k$. In our model, the number of subgroups $K$ can later be determined in a data-driven manner through the use of information criteria.\footnote{This is different from many other clustering techniques, such as $k$-means clustering, where heuristics are employed to determine the number of clusters $K$.}

In our \model, each subgroup is associated with its own HMM (and thus a separate parameterization), which allows us to model the disease progression in each subgroup differently. Specifically, each subgroup has a different hidden Markov model with a subgroup-specific parameterization, $m^k$ with $k=1,\ldots,K$. In this sense, the subgroup probability $\omega_i^k$ matches a patient $i$ to one of the HMM parameterizations $m^k$. 

The variable $\omega_i^k$ denotes the probability of a patient belonging to a subgroup and thus needs to be estimated from the data. That is, the probability follows the logic that it links a patient to the most probable progression model, \ie, $\omega_i^k = P(m^k \mid x_i)$. This variable is now rewritten such that it incorporates the structural information from the risk factors. We thus model the variable $\omega_i^k$ as a multinomial logit model
\begin{equation}\label{eq:weights}
\omega_i^k = P(m^k \mid x_i) = \frac{\exp(\alpha^k + x_i^T \beta^k)}{\sum_{j=1}^{K} \exp(\alpha^j + x_i^T \beta^j)} 
\end{equation}
with coefficient vectors $\beta^k$ for the risk factors and intercepts $\alpha^k$ across all subgroups $k = 1, \ldots, K$.

The above multinomial logit model from Eq.~\eqref{eq:weights} is interpretable to help generate a better understanding of which risk factors are indicative of a specific subgroup and thus which risk factors explain specific disease dynamics. To this end, the coefficients $\beta^k$ yield disease markers that are characteristic of a specific subform of a disease. Such disease markers may then enable the identification -- and, subsequently, the customization of treatments -- for a specific subgroup \citep{strimbu2010biomarkers}.

\subsubsection{Subgroup-specific HMMs}

Each HMM $m^k$, $k = 1, \ldots, K$, consists of three components, namely, (1)~latent states, (2)~a transition component, and (3)~an emission component linking latent states and observations. These are as follows:

\emph{(1) Latent states.} Each HMM introduces latent states $s_{it} \in \{1, \ldots, \mathcal{S} \}$ for patient $i$ and time steps $t = 1, \ldots, T$. The latent states themselves cannot be directly observed, but are stochastically linked with the observations, \ie, the symptoms $y_{it}$, through the emission component. %

We set the number of latent states to $\mathcal{S} = 3$, as stated in the trajectory framework \citep{Corbin.1991,Corbin.1998}. In Supplement~\ref{appendix:robustness_states}, we provide further justification. Therein, we perform an empirical analysis where we compare different choices of $\mathcal{S}$ and find that $\mathcal{S} = 3$ gives the best fit, which justifies our choice throughout the rest of the paper.  

\emph{(2) Transitions.} The succession of latent states is described by a transition probability $\phi^k_{rs}$ that denotes the probability of a patient moving from latent state $r$ at $t$ to latent state $s$ at $t+1$. This is denoted by $P(s_{i,t+1} = s\mid s_{it}=r)$. The initial state distribution at $t = 1$ is given $\pi^k_s = P(s_{i1} = s)$. The transition probabilities then form a matrix $\Phi^k$ that is later estimated from data.

\emph{(3) Emissions.} The emission component links the latent states and the observations. For this, we introduce the so-called emission probability $b^k_{s}(y_{it}) = P(y_{it}\mid s_{it}=s)$ for a given state $s=1,\ldots,\mathcal{S}$. Hence, observations are modeled as being dependent on the latent state. As is common for disease monitoring, we have multivariate observations, which we accommodate via a copula approach (see the following section). That is, $b^k_{s}$ essentially returns the joint multivariate distribution across multiple symptoms. 

Altogether, each hidden Markov model $m^k$ is parameterized by a set of initial state distributions $\pi^k$, a transition matrix $\Phi^k$, and an emission component $b^k$.

\subsubsection{Copula approach for multivariate discrete observations}
\label{sec:copula}

Our \model is fitted to a multivariate time series with multiple symptoms. However, symptoms are typically not independent; rather, they co-occur in a systematic manner. For example, the absence of pain frequently co-occurs with the absence of disability. Hence, we design our \model to incorporate multivariate observations with an explicit dependence structure through a copula approach. Recall that $b^k_{s}$ essentially returns the joint multivariate distribution across multiple symptoms. We now formalize $b^k_{s}$ by modeling the marginal distributions separately but, to account for the dependence, introduce an additional parameterized copula.

We assume the joint distribution of the variables to be expressed by a subgroup-specific copula $C_{\rho^k}(u,v)$, \ie, a function $C_{\rho^k}\colon[0,1]^2\to[0,1]$ for all subgroups $k = 1, \ldots, K$.\footnote{Let $C(\cdot; \ldots ; \cdot)$ denote a copula function. A copula \citep{Joe.1997} is a generalized correlation function, which allows for a stronger dependence in certain parts of the distribution. Mathematically, it links the individual marginal distributions of a multi-dimensional input to a joint cumulative distribution function, while introducing a desired dependence structure between the margins. Depending on which copula is chosen, $C$ might be parameterized by further variables that, for instance, control for the strength of the dependence. Here, we allow the copula to depend on a subgroup-specific parameter $\rho^k$, $k=1, \ldots, K$.} Notably, the copula parameter $\rho^k$ is also allowed to vary across the different subgroups $k = 1, \ldots, K$. Then, the joint distribution function $F$ of $y_{it}^P$ and $y_{it}^D$ can be rewritten as
\begin{equation}
F \left( y_{it}^P, y_{it}^D \right) = C_{\rho^k} \left( F_P \left( y_{it}^P \right), F_D \left( y_{it}^D \right) \right),
\end{equation}
where the cumulative distribution function of the margin $y_{it}^P$ and $y_{it}^D$ are given by $F_P$ and $F_D$, respectively. 

In our analyses, we decided upon a survival Gumbel copula, which is given by
{\small\begin{multline*}
C_{\rho^k}(u,v) = u + v -1 \\
+ \exp\left[-\left( (-\log{1-u})^{\rho^k} + (-\log{1-v})^{\rho^k} \right)^{\frac{1}{\rho^k}}\right]
\end{multline*}}
for $u,v\in(0,1)$ and $\rho^k \geq1$. The advantage of this choice is that it allows us to model a stronger lower tail dependence.\footnote{It can be shown that the survival Gumbel copula has positive lower tail dependence for $\nu_s > 1$ and zero upper tail dependence for all $\nu_s$
\citep{Joe.1997} For the special case of $\nu_s = 1$, the survival Gumbel copula reduces to independent observations.} The reasons are the following: 
\begin{itemize}
\item \emph{Theoretical justification:} Symptoms co-occur in a specific manner: (1)~either \emph{all} (or almost all) symptoms are absent when the patient has recovered, or (2)~the condition is indicated by the presence \emph{some} symptoms. For instance, low back patients suffer typically from either/or pain/disability but rarely both. To address this mathematically, the absence of one characteristic must make it more likely that other characteristics will also be absent. Altogether, this results in a lower tail dependence among health measurements that must be modeled accordingly. The survival Gumbel copula has exactly this behavior. 
\item \emph{Empirical justification:} In our analyses, we tested a variety of copulas. Specifically, we used the VineCopula package for R to compare a range of alternative copulas numerically, namely, tawn type II, BB7, Fran, and Joe, but found that the survival Gumbel copula gave the best empirical fit. 
\end{itemize}
More details behind our choice are in Supplement~\ref{appendix:robustness_states}.

Using the above copula, we introduce a dependence structure between the marginal distribution of $b_s(y_{it} \leq y)$. Here, we yield
{\small\begin{align*}
& b_s(y_{it} \leq y) = P(y_{it}^{P}\leq y_P, y_{it}^{D}\leq y_D\mid s_{it}=s) \\
 = & C_{s}\left(P(y_{it}^{P}\leq y_P\mid s_{it}=s),P(y_{it}^{D}\leq y_D\mid s_{it}=s)\right) .
\end{align*}}%
As a result, the dependence between observations is modeled by \textquote{linking} the separate marginal distribution functions $P(y_{it}^{P}\leq y_P\mid s_{it}=s)$, $P(y_{it}^{D}\leq y_D\mid s_{it}=s)$ inside the copula \citep[\eg,][]{Nikoloulopoulos.2013}. Based on the distribution function $b_s(y_{it} \leq y)$, we now derive the new emission $b_s(y_{it}=y)$ via
{\small\begin{align}
&b_s(y_{it} = y) =P(y_{it}^{P} = y_P, y_{it}^{D} = y_D\mid s_{it} = s)\\
=& \sum_{i_P=0}^1 \sum_{i_D=0}^1(-1)^{i_P+i_D} \, P({y}_{it}^{P}\leq {y}_P-i_P,  \nonumber \\ 
& \qquad\qquad\qquad\qquad y_{it}^{D}\leq {y}_D-i_D\mid s_{it}=s) \label{eqn:emission_derivation_sum} \\
=&\sum_{i_P=0}^1 \sum_{i_D=0}^1(-1)^{i_P+i_D} \, C_{\nu_s}\Big(P({y}_{it}^{P}\leq {y}_P-i_P\mid s_{it}=s), \nonumber \\ 
& \qquad\qquad\qquad P(y_{it}^{D}\leq y_D-i_D\mid {s}_{it}=s)\Big) . 
 \label{eqn:emission_derivation_copula}
\end{align}}%
The equation follows from the fact that the marginal likelihood can be derived from the corresponding distribution function as $b_s^{\kappa}(y_{it}^{\kappa}=y_\kappa) = P(y_{it}^{\kappa}\leq y_\kappa\mid s_{it}=s) - P(y_{it}^{\kappa}\leq y_\kappa\mid s_{it}=s)$, $\kappa \in \{ P, D \}$. Eq.~\eqref{eqn:emission_derivation_copula} is thus the result of inserting the distribution function from above.

\subsection{Model estimation}
\label{sec:estimation_procedure}

We estimate our \model using a so-called ``fully'' Bayesian approach \citep{Gelman.2014}. Details are in Supplement~\ref{appendix:model_estimation}. Specifically, we use Markov chain Monte Carlo in order to sample from the joint posterior distribution of the model parameters, and, for this, we derive the likelihood $L$ for our \model.

\begin{figure}
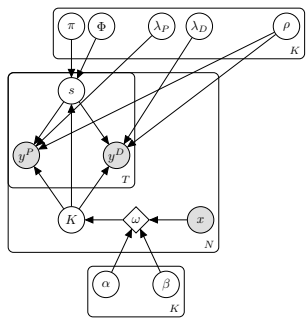

	\centering
 \vspace{-0.2cm}
        \scalebox{0.5}{
	\tikz{
		\node[latent](State){$s$};

		\node[obs,below=of State,xshift=-1.2cm] (Pain) {$y^P$};%
		\node[obs,below=of State,xshift=1.2cm] (Disability) {$y^D$};%
		\node[latent,below=of Pain,xshift=1.2cm](Cluster){$K$};
		\node[det,right=of Cluster](Weight){$\omega$};
		\node[obs,right=of Weight](Risk){$x$};
		\node[latent,below=of Weight,xshift=-0.8cm](Inter){$\alpha$};
		\node[latent,below=of Weight,xshift=0.8cm](Slope){$\beta$};
		\node[latent,above=of State](Init){$\pi$};
		\node[latent,above=of State,xshift=0.8cm](Tpm){$\Phi$};
		\node[latent,above=of State,xshift=2.4cm](Painemis){$\lambda_P$};
		\node[latent,above=of State,xshift=3.4cm](Disemis){$\lambda_D$};
		\node[latent,right=of Disemis,xshift=0.6cm](Corr){$\rho$};		
		\plate [] {plate1} {(Pain)(Disability)(State)} {$T$}; %
		\plate [] {plate2} {(Pain)(Disability)(State)(Cluster)(Weight)(Risk)} {$N$};
		\plate [] {plate3} {(Slope)(Inter)} {$K$};
		\plate [] {plate4} {(Init)(Tpm)(Painemis)(Disemis)(Corr)} {$K$};
		\edge {State}{Pain};
		\edge {Painemis,Corr}{Pain};
		\edge {Disemis,Corr}{Disability};
		\edge {State}{Disability};
		\edge {Cluster}{State,Pain,Disability};
		\edge {Init,Tpm}{State};
		\edge {Weight}{Cluster};
		\edge {Inter,Slope}{Weight};
		\edge {Risk}{Weight}}
  }
\vspace{-0.2cm}
\caption{Plate notation.}
\label{fig:plate_notation}
  \vspace{-.5cm}
\end{figure}

A summary of the model in plate notation is given in Fig.~\ref{fig:plate_notation}.

\textbf{Likelihood:} The posterior distribution is computed using the \mbox{(log-)}likelihood $L$, which is derived from the likelihood corresponding to each subgroup-specific HMM weighted by $\omega_i^1,\ldots,\omega_i^K$. Mathematically, we derive
{\small%
\begin{align*}\label{eq:mhmm_ll}
& L = \sum_{i=1}^N\log{P \bigg( \underbrace{ y_{i1}^P,y_{i1}^D,\ldots,y_{iT}^P, y_{iT}^D }_{T \textup{ time steps}} \bigg| x_i \bigg)}\\
& = \sum_{i=1}^{N}\log\left[\sum_{k=1}^KP\left(m^k \middle| x_i\right)\, P \left(y_{i1}^P,y_{i1}^D,\ldots\middle| m^k \right)\right]\\ 
&= \sum_{i=1}^{N} \log\left[\sum_{k=1}^{K} \omega_i^k \, P \left(y_{i1}^P,y_{i1}^D,\ldots,y_{iT}^P ,y_{iT}^D \middle| m^k \right)\right] , 
\end{align*}}
where we highlighted how the time series of pain and disability enters the estimation procedure.

\subsection{Patient assignment to subgroups}
\label{sec:subgroup_assignment}

Once a set of subgroups together with the weights $\omega_i^k$ has been determined from historical data, new patients can be assigned to these subgroups. Here, we essentially distinguish two cases:\footnote{We use the term offline/online as in online algorithms where data is processed piece-by-piece (and not in the sense of reinforcement learning).} (1)~If an incoming patient arrives for the first time at a care provider, the initial step is to file a diagnosis of the condition. At this point in time, little is known about the specific subtype of the disease, since its progression was not monitored before. Hence, one can merely draw upon patient-specific variables $x_i$. We refer to this as the \textquote{offline} assignment. (2)~Once the patient has been subject to monitoring, one can generate more extensive knowledge of how symptoms evolve. Hence, one has additionally access to a time series with symptom $y_{it}$, which can then be leveraged to assign patients to subgroups. We refer to this as \textquote{online} assignment.  

In the following, we denote the assignment by $G_{it}$ for a given patient $i$ and where the first $t$ time steps from the trajectory have been observed. Accordingly, we distinguish two different strategies, namely, offline and online assignments. Formally, the offline setting refers to cases where incoming patients, for whom no trajectory has been collected (\ie, $t = 0$), are matched to a subgroup. Hence, the assignment merely draws upon the risk factors $x_i$. In contrast, the online setting additionally includes time series data (\ie, $t \neq 0$) and can also be utilized during assignment. Both are formalized in the following (see Table~\ref{tbl:subgroup_assignment}):

\emph{Offline assignment ($t = 0$).} By using the estimated coefficients $\alpha^k$ and $\beta^k$, one can compute the corresponding subgroup probabilities $\omega_i^k$ for an incoming patient $i$ and compare them across all subgroups $k = 1, \ldots, K$. The final assignment $G_{i0}$ is then based upon the maximum probability, given by 
\begin{equation}
G_{i0}=\argmax_{k = 1, \ldots, K} P\left(m^k \middle| x_i\right) = \argmax_{k=1,\ldots,K}\omega_i^k.
\end{equation}
By definition, the formula for $\omega_i^k$ merely refers to the risk factors $x_i$ and matches a patient to the most probable disease progression model; hence, it does not need actual observations from the health trajectory. 

\emph{Online assignment ($t \neq 0$).} In contrast to the static offline subgroup assignment, the online subgroup assignment can be dynamically updated for arbitrary $t$ once the corresponding window of the health trajectory is available. This approach aims at choosing the subgroup that maximizes the likelihood of the observed trajectory conditional on the risk factors. It includes the likelihood of the observations $y_{i1}^P,y_{i1}^D,\ldots,y_{it}^P,y_{it}^D$ under the model $m^k$. The online assignment is thus given by
\vspace{-0.4cm}
\begin{align*}
& G_{it} = \argmax_{k=1,\ldots, K}P\left(m^k \middle| y_{i1}^P,y_{i1}^D,\ldots y_{it}^P,y_{it}^D,x_i \right) \\ 
&= \argmax_{k = 1, \ldots, K} P\left(m^k \middle| x_i\right) P\left(y_{i1}^P,y_{i1}^D,\ldots y_{it}^P,y_{it}^D \middle| m^k\right)\\
&= \argmax_{k=1,\ldots,K}\omega_i^k\,P\left(y_{i1}^P,y_{i1}^D,\ldots y_{it}^P,y_{it}^D \middle| m^k\right) .
\end{align*}

\begin{table*}[h]
	\centering
	\tiny
	\begin{tabular}{lp{3cm}lp{6cm}}
		\toprule
		Assignment & Input & Subgroup probability & Application  \\
		\midrule
		Offline & Patient-specific risk factors $x_i$ & $\omega_i^k$  &Incoming patients with freshly-diagnosed condition for whom no symptoms have yet been collected \\
		\midrule
		Online & Patient-specific risk factors $x_i$ and time series data $y_{it}$ with symptoms & $\omega_i^k\,P(y_{i1}^P,y_{i1}^D,\ldots y_{it}^P,y_{it}^D\mid m^k)$ & Patients with previously observed progression of symptoms for whom the trajectory can be leveraged to determine the subgroup assignment \\
		\bottomrule
	\end{tabular}
 \vspace{-0.2cm}
	\caption{Comparison of online and offline approaches to subgroup assignments.}
	\label{tbl:subgroup_assignment}
 \vspace{-0.3cm}
\end{table*}

\vspace{-.5cm}
\section{Data}
\label{sec:setting}

Non-specific low back pain is typically a chronic condition and globally responsible for the greatest number of years lived with disability \citep{GBD.2019}. Low back pain is an ideal setting to demonstrate our model: Afflicted patients are known for their considerable heterogeneity, which is so far understood only poorly \citep{Silva.2022} and for which the specific causes still remain largely unknown \citep{Hartvigsen.2018}. Hence, this makes low back pain a compelling  case of direct clinical relevance where the identification of novel disease markers fulfills a direct need in practice.

Our clinical study lasted for 52 weeks and counted 847 patients. This exceeds the usual length of low back pain episodes by several orders \citep{Kongsted.2015}. In our study, \SI{62.54}{\percent} of all low back pain episodes last up to 2~weeks, \SI{13.69}{\percent} between 2--4 weeks and \SI{10.68}{\percent} between 1--3 months. Hence, a length of 52 weeks for our longitudinal study is sufficient to capture multiple episodes with (severe) low back pain.

For each patient, our dataset includes: (1)~A comprehensive set of baseline variables with potential risk factors $x_i$ (\eg, age, height, BMI). (2)~The progression of low back pain with $T=52$ weekly observations $y_{it}^P$ and $y_{it}^D$ for pain intensity and days with activity limitation, respectively. (3)~Outcomes from a follow-up after the clinical study, that is, after 12 months. Details are in Supplement~\ref{appendix:dataset}.

\vspace{-0.3cm}
\section{Empirical results}
\label{sec:results}

\subsection{Model fit}

We followed the procedure in \citep{Gelman.2014} to perform model selection, that is, to determine the number of subgroups $K$ and the number of latent states $\mathcal{S}$. Our estimation returns $K = 8$. Details for this are in Supplement~\ref{appendix:estimation_model}. Analogously, our empirical results determined $\mathcal{S} = 3$ latent states. The results are in Supplement~\ref{appendix:robustness_states}.

\vspace{-0.3cm}
\subsection{Benchmarking} 
\label{sec:subgroup_comparison}

\textbf{Baselines:} In health management, it is common to perform subgrouping only based on risk factors and, therefore, without information on health trajectories \citep[\eg,][]{Nielsen.2016}. We compare our proposed \model for subgrouping against baselines: (1)~We use a latent class analysis (LCA) that performs clustering based on the raw, single-item values for all risk factors \citep{Nielsen.2016}. (2)~We use a two-staged variant where the risk factors are first grouped into meaningful categories (physical activity, pain intensity, work/social activities, physical impairment, psychological state and contextual factors). Then, a summary score is computed for the variables from each category. This was done using established procedures from validated questionnaires in the medical domain to map individual scores onto an overall score \citep{Nielsen.2016}. Afterward, the summary scores for the different categories were inserted into latent class analysis. (3)~We perform clustering via the $k$-means algorithm using the different risk factors \citep[cf.][]{Hastie.2009}. Here, we follow the standard approach of choosing an appropriate number of clusters by manually examining the relative within-cluster sum of squares and obtain 8 subgroups. Notably, for all three baselines, the same risk factors as in our \model are used.\footnote{We also experimented with other approaches where we accommodated the time-series data from the health trajectory in addition to the baseline risk factors as input to the $k$-means algorithm but found it to be inferior. This has intuitive reasons: First, this would not allow to assignment of incoming patients without trajectory data to one of the subgroups. Further, let us assume two patients who both experience alternating patterns of episodes with high and low pain. Calculating the L2-norm as is done in $k$-means would treat both health trajectories as highly dissimilar, even though both have the same patterns but simply with a different offset. As seen in this example, a model-based approach for time-series clustering has clear advantages over a model-free approach, especially for volatile data such as health trajectories with symptoms.} (4)~We experimented with hierarchical clustering. We omitted it for space as it was similar to $k$-means.

\begin{table*}[htbp]
\centering
		\tiny
		\begin{tabular}{lc SSS SSS}
			\toprule
			&&\multicolumn{3}{c}{Panel A: Pain} & \multicolumn{3}{c}{Panel B: Activity limitation}\\
			\cmidrule(lr){3-5} \cmidrule(lr){6-8} 
			Clustering approach & Subgroups & \multicolumn{1}{c}{Sil$\uparrow$} & \multicolumn{1}{c}{CH$\uparrow$} & \multicolumn{1}{c}{DB*$\downarrow$} & \multicolumn{1}{c}{Sil$\uparrow$} & \multicolumn{1}{c}{CH$\uparrow$} & \multicolumn{1}{c}{DB*$\downarrow$} \\
			\midrule
			\csname @@input\endcsname CVI_comparison.tex
			\bottomrule
    \multicolumn{8}{l}{Best value in bold. Stated : Silhouette~(Sil$\uparrow$), Calinski-Harabasz~(CH$\uparrow$) and Davies-Bouldin~(DB*$\downarrow$) indices.}
		\end{tabular}
  \vspace{-0.2cm}
	\caption{Comparison of subgrouping for observed trajectories in the test set.}
	\label{tbl:cluster_comparison}
   \vspace{-0.4cm}
\end{table*}

\vspace{-0.3cm}
\subsection{Performance comparison}

\textbf{Performance metrics:} We compare the clustering performance in terms of so-called cluster validity indices~\citep{Arbelaitz.2013}. Specifically, we use: \textbf{Calinski-Harabasz}~(CH$\uparrow$), \textbf{Silhouette}~(Sil$\uparrow$), and \textbf{Davies-Bouldin}~(DB*$\downarrow$). The arrow is used to indicate whether larger or smaller values are preferred. Details are in Supplement~\ref{appendix:cluster_validity_indices}. Reporting is done separately for pain and disability to allow for more granular insights.


\textbf{Results:} Our proposed \model subgrouping scores best for five out of six metrics (Table~\ref{tbl:cluster_comparison}). For instance, our subgrouping approach improves the Davies-Bouldin index (DB*$\downarrow$) over the best latent class analysis by \SI{32.9}{\percent} (for pain) and by \SI{29.9}{\percent} (for activity limitation).  Further, we can quantify the additional improvement in terms of cluster validity indices resulting from the use of health trajectories. For example, the Calinski-Harabasz~(CH$\uparrow$) index of the latent class analysis is \num{4.6} times inferior to the one from our \model; similarly, the Davies-Bouldin~(DB*$\downarrow$) index scores \SI{29.9}{\percent} better in the case of our \model. Overall, the latent class analysis is never superior to our \model, with the exception of the Silhouette index (Sil$\uparrow$) index for activity limitation. 


\textbf{Comparison to baselines:} We also experimented with other baselines (not shown for reasons of space). These generally require access to the longitudinal information for clustering, which may help in modeling the data but such baselines are impractical in clinical use where incoming patients must be assigned to subgroups without having access longitudinal information. First, we experimented with recurrent neural networks to describe patient trajectories such as LSTMs. However, LSTMs are inferior to HMM-based approaches in modeling low back pain trajectories \citep{EJOR}, because of which the subgrouping is also inferior. Second, we considered other neural approaches such as SOM-VAE \citep{Fortuin.2019} and T-DPSOM \citep{Manduchi.2021}. Overparameterized models such as SOM-VAE and T-DPSOM can have advantages for large-scale clinical registries, while our model is especially suited for small sample sizes for real-world clinical studies as ours. Further, neural approaches capture the underlying latent state dynamics, which is another explanation as to why our model is superior.  

Finally, group-based trajectory modeling \citep{Nagin2024} can be seen as a simplified version of our MHMMX (i.e., using a single state, which turns the latent dynamics into a Markov chain). We evaluated a single-state MHMMX as part of our ablations (Table~\ref{tbl:mhmm_cluster_states}), finding that it gives an inferior fit. 

\subsection{Interpretation of subgroups}
\label{sec:results_subgroups_interpretation}

\begin{figure*}[h]
	\centering
	\includegraphics[width=.65\linewidth]{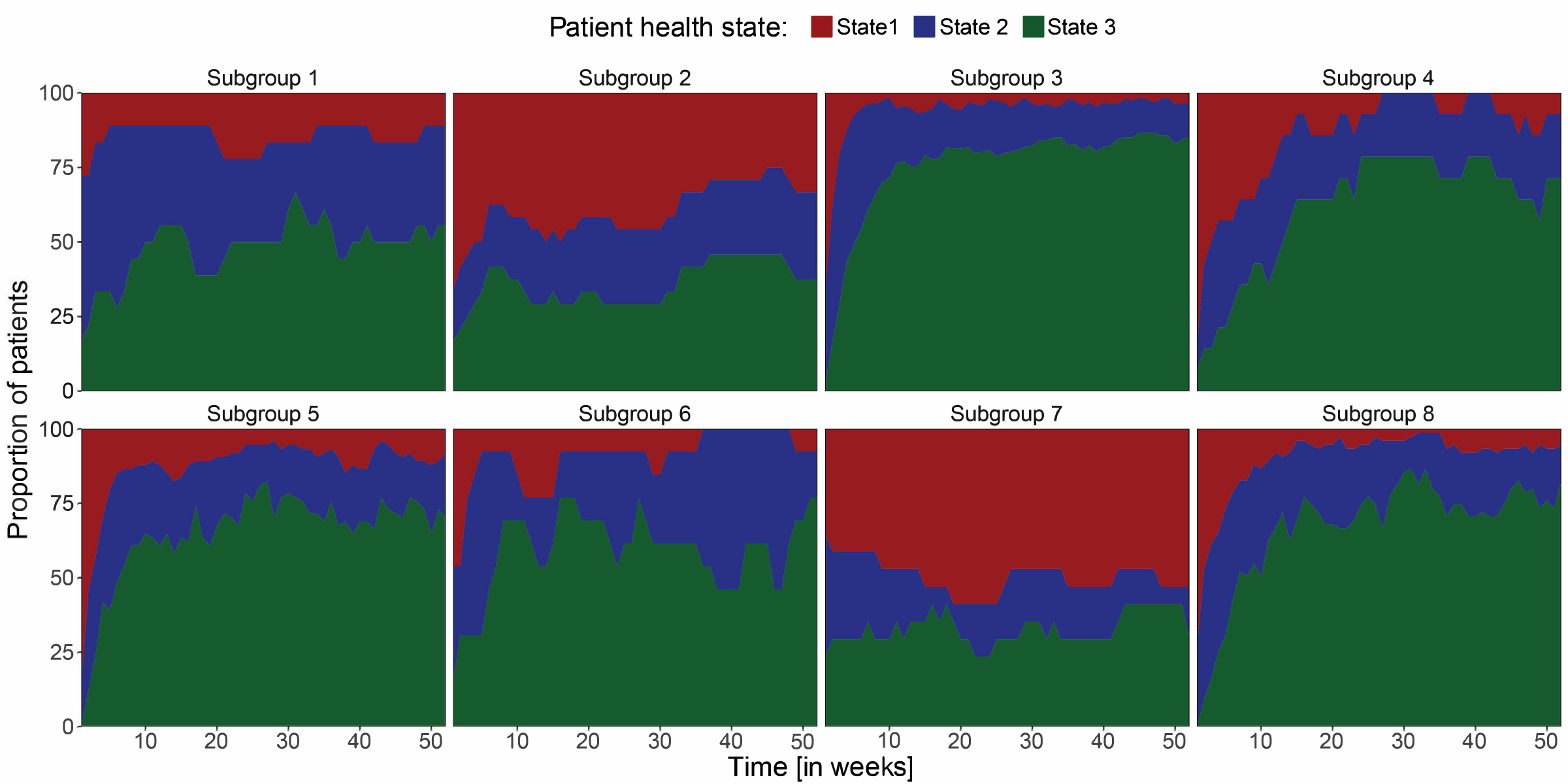}
	\caption{Comparison of the recovered latent states over time. Each plot compares a different subgroup from our \model and reports the relative proportion of patients being in each of the latent states at a given point in time, thereby demonstrating considerable differences in the disease progression across the different subgroups. Here, states can be interpreted as follows: state~1 (``severe''), state~2 (``moderate''), and state 3 (``mild''). }
	\label{fig:patient_cluster}
 \vspace{-1cm}
\end{figure*}

One of the authors (AK)  with expertise in the diagnosis and treatment of low back pain and a clinical researcher with expertise in low back pain subgrouping were involved to discuss how the different subgroups obtained by our model lend to potentially clinically-relevant interpretation. The results are in Supplement~\ref{appendix:extended_results}. For example, our \model identifies 8 subgroups among patients. One subgroup experiences alternating phases of recovery and severe symptoms, whereas the trajectories of another show mostly relief from pain and disability after 12 months (see Fig.~\ref{fig:patient_cluster}). For example, our health professionals describe subgroup 3 as ``recovery'', while subgroup 2 was characterized as ``severe''. Hence, each subgroup should receive a different treatment plan. As such, our model generates new insights that may prove meaningful for disease management and that may help in identifying new disease markers. 

\vspace{-0.3cm}
\subsection{Validation through clinical experts}

We worked with health professionals with expertise in the diagnosis and treatment of low back pain to discuss how the different subgroups obtained by our model lend to potentially clinically-relevant interpretation. Our validation is inspired by the objectives in clustering that should lead to (a) large dissimilarities in between-subgroup comparisons and (b) large similarities in within-subgroup evaluations. According to our clinical researchers, both criteria are relevant because they ensure that practitioners can match patients to specific subgroups in day-to-day care as well as it may inform   tailored treatment plans for each subgroup, with the intent to improve treatment outcomes. 

We provided information about trajectory dynamics to the clinical researchers as well as overall baseline characteristics for each subgroup. We were specifically interested in (a) the between-subgroup comparisons reveal sufficient dissimilarities and whether (b) the within-subgroup trajectories are of sufficient similarities to warrant a joint subgroup. Both led to positive evaluations. Hence, we concluded based on our expert validation that subgrouping yields subgroups that are clinically-interpretable as desired.

\vspace{-0.3cm}
\subsection{Validation through post-hoc outcomes}

Finally, we use our follow-up survey after the 12 months to assess whether subgroups have different outcomes in terms of pain and disability. The results are in Fig.~\ref{tbl:interpretation_clusters}, confirming that our subgroups can successfully capture heterogeneity in outcomes. For example, we find that subgroup 3 (``recovery'') has, on average, very low pain and almost no disability, while subgroup 2 (``severe'') has larger pain and disability by several orders.

\vspace{-0.3cm}
\subsection{Comparison of offline and online subgroup assignment}

Our proposed approach allows for both offline and online subgroup assignment of patients. A detailed comparison is in Supplement~\ref{appendix:online_offline_assignment}. The results demonstrate that our framework is effective for subgroup assignments of both (1)~incoming patients without longitudinal information and (2)~patients with longitudinal information.

\vspace{-0.3cm}
\section{Discussion}
\label{sec:discussion}

The clear benefit of our \model in clinical practice is the ability to identify subgroups that are informative of the future trajectory pattern as it unfolds. As such the purpose of our \model is foremost about generating insights for understanding and explaining.  
Finally, our model is not only applicable to low back pain but also to other chronic conditions or conditions likely to become chronic and thus follow the so-called trajectory framework.



\vspace{0.2cm}
{\footnotesize\textbf{Acknowledgments.} Financial supported from the Swiss National Science Foundation (SNSF) via grant 186932 is acknowledged.}

\FloatBarrier

\bibliography{literature}

\begin{thebibliography}{61}
\providecommand{\natexlab}[1]{#1}
\providecommand{\url}[1]{\texttt{#1}}
\expandafter\ifx\csname urlstyle\endcsname\relax
  \providecommand{\doi}[1]{doi: #1}\else
  \providecommand{\doi}{doi: \begingroup \urlstyle{rm}\Url}\fi

\bibitem[Aghabozorgi et~al.(2015)Aghabozorgi, {Seyed Shirkhorshidi}, and {Ying
  Wah}]{Aghabozorgi.2015}
Saeed Aghabozorgi, Ali {Seyed Shirkhorshidi}, and Teh {Ying Wah}.
\newblock Time-series clustering: A decade review.
\newblock \emph{Information Systems}, 53:\penalty0 16--38, 2015.

\bibitem[Allam et~al.(2021)Allam, Feuerriegel, Rebhan, and
  Krauthammer]{Allam.2021}
Ahmed Allam, Stefan Feuerriegel, Michael Rebhan, and Michael Krauthammer.
\newblock Analyzing patient trajectories with artificial intelligence.
\newblock \emph{Journal of Medical Internet Research}, 23\penalty0
  (12):\penalty0 e29812, 2021.

\bibitem[Arbelaitz et~al.(2013)Arbelaitz, Gurrutxaga, Muguerza, P{\'e}rez, and
  Perona]{Arbelaitz.2013}
Olatz Arbelaitz, Ibai Gurrutxaga, Javier Muguerza, Jes{\'u}s~M. P{\'e}rez, and
  I{\~n}igo Perona.
\newblock An extensive comparative study of cluster validity indices.
\newblock \emph{Pattern Recognition}, 46\penalty0 (1):\penalty0 243--256, 2013.

\bibitem[Ashley(2015)]{Ashley.2015}
Euan~A. Ashley.
\newblock The precision medicine initiative: A new national effort.
\newblock \emph{JAMA}, 313\penalty0 (21):\penalty0 2119--2120, 2015.

\bibitem[Bertsimas et~al.(2017)Bertsimas, Kallus, Weinstein, and
  Zhuo]{Bertsimas.2017}
Dimitris Bertsimas, Nathan Kallus, Alexander~M. Weinstein, and Ying~Daisy Zhuo.
\newblock Personalized diabetes management using electronic medical records.
\newblock \emph{Diabetes Care}, 40\penalty0 (2):\penalty0 210--217, 2017.

\bibitem[Breivik et~al.(2008)Breivik, Borchgrevink, Allen, Rosseland,
  Romundstad, Hals, Kvarstein, and Stubhaug]{Breivik.2008}
Harald Breivik, P.~C. Borchgrevink, S.~M. Allen, Leiv~A. Rosseland, Liv
  Romundstad, E.~K.~Breivik Hals, Gunnvald Kvarstein, and Audun Stubhaug.
\newblock Assessment of pain.
\newblock \emph{British Journal of Anaesthesia}, 101\penalty0 (1):\penalty0
  17--24, 2008.

\bibitem[Cohen et~al.(2010)Cohen, Grossman, Morabito, Knudson, Butte, and
  Manley]{Cohen.2010}
Mitchell~J. Cohen, Adam~D. Grossman, Diane Morabito, M.~Margaret Knudson,
  Atul~J. Butte, and Geoffrey~T. Manley.
\newblock Identification of complex metabolic states in critically injured
  patients using bioinformatic cluster analysis.
\newblock \emph{Critical Care}, 14\penalty0 (1):\penalty0 R10, 2010.

\bibitem[Corbin(1998)]{Corbin.1998}
Juliet Corbin.
\newblock The {C}orbin and {S}trauss chronic illness trajectory model: An
  update.
\newblock \emph{Scholarly Inquiry for Nursing Practice}, 12\penalty0
  (1):\penalty0 33--41, 1998.

\bibitem[Corbin and Strauss(1991)]{Corbin.1991}
Juliet Corbin and Anselm Strauss.
\newblock A nursing model for chronic illness management based upon the
  trajectory framework.
\newblock \emph{Scholarly Inquiry for Nursing Practice}, 5\penalty0
  (3):\penalty0 155--174, 1991.

\bibitem[Corbin and Strauss(1988)]{Corbin.1988}
Juliet~M. Corbin and Anselm Strauss.
\newblock \emph{Unending work and care: Managing chronic illness at home}.
\newblock Jossey-Bass, 1988.
\newblock ISBN 1555420826.

\bibitem[DeSantis and Bandyopadhyay(2011)]{DeSantis.2011}
Stacia~M. DeSantis and Dipankar Bandyopadhyay.
\newblock Hidden {M}arkov models for zero-inflated {P}oisson counts with an
  application to substance use.
\newblock \emph{Statistics in Medicine}, 30\penalty0 (14):\penalty0 1678--1694,
  2011.

\bibitem[Downie et~al.(2016)Downie, Hancock, Rzewuska, Williams, Lin, and
  Maher]{Downie.2016}
Aron~S. Downie, Mark~J. Hancock, Magdalena Rzewuska, Christopher~M. Williams,
  Chung-Wei~Christine Lin, and Christopher~G. Maher.
\newblock Trajectories of acute low back pain: A latent class growth analysis.
\newblock \emph{Pain}, 157\penalty0 (1):\penalty0 225--234, 2016.

\bibitem[Feuerriegel et~al.(2024)Feuerriegel, Frauen, Melnychuk, Schweisthal,
  Hess, Curth, Bauer, Kilbertus, Kohane, and {van der
  Schaar}]{Feuerriegel.2024}
Stefan Feuerriegel, Dennis Frauen, Valentyn Melnychuk, Jonas Schweisthal,
  Konstantin Hess, Alicia Curth, Stefan Bauer, Niki Kilbertus, Isaac~S. Kohane,
  and Mihaela {van der Schaar}.
\newblock Causal machine learning for predicting treatment outcomes.
\newblock \emph{{Nature Medicine}}, 2024.

\bibitem[Fortuin et~al.(2019)Fortuin, H{\"u}ser, Locatello, Strathmann, and
  R{\"a}tsch]{Fortuin.2019}
Vincent Fortuin, Matthias H{\"u}ser, Francesco Locatello, Heiko Strathmann, and
  Gunnar R{\"a}tsch.
\newblock {SOM-VAE}: Interpretable discrete representation learning on time
  series.
\newblock In \emph{International Conference on Learning Representations}, 2019.

\bibitem[Gelman et~al.(2014)Gelman, Hwang, and Vehtari]{Gelman.2014}
Andrew Gelman, Jessica Hwang, and Aki Vehtari.
\newblock Understanding predictive information criteria for {B}ayesian models.
\newblock \emph{Statistics and Computing}, 24\penalty0 (6):\penalty0 997--1016,
  2014.

\bibitem[Ghassempour et~al.(2014)Ghassempour, Girosi, and
  Maeder]{Ghassempour.2014}
Shima Ghassempour, Federico Girosi, and Anthony Maeder.
\newblock Clustering multivariate time series using hidden {M}arkov models.
\newblock \emph{International Journal of Environmental Research and Public
  Health}, 11\penalty0 (3):\penalty0 2741--2763, 2014.

\bibitem[Gon{\c{c}}alves et~al.(2023)Gon{\c{c}}alves, Poyraz, Paul, and
  Lawn]{gonccalves2023inferring}
Bronner~P Gon{\c{c}}alves, Onur Poyraz, Proma Paul, and Joy~E Lawn.
\newblock Inferring longitudinal patterns of group b {S}treptococcus
  colonization during pregnancy.
\newblock \emph{iScience}, 26\penalty0 (7), 2023.

\bibitem[Hamburg and Collins(2010)]{Hamburg.2010}
Margaret~A. Hamburg and Francis~S. Collins.
\newblock The path to personalized medicine.
\newblock \emph{New England Journal of Medicine}, 363\penalty0 (4):\penalty0
  301--304, 2010.

\bibitem[Hartvigsen et~al.(2018)Hartvigsen, Hancock, Kongsted, Louw, Ferreira,
  Genevay, Hoy, Karppinen, Pransky, Sieper, Smeets, Underwood, Buchbinder,
  Cherkin, Foster, Maher, {van Tulder}, Anema, Chou, Cohen, {Menezes Costa},
  Croft, Ferreira, Ferreira, Fritz, Gross, Koes, {\"O}berg, Peul, Schoene,
  Turner, and Woolf]{Hartvigsen.2018}
Jan Hartvigsen, Mark~J. Hancock, Alice Kongsted, Quinette Louw, Manuela~L.
  Ferreira, St{\'e}phane Genevay, Damian Hoy, Jaro Karppinen, Glenn Pransky,
  Joachim Sieper, Rob~J. Smeets, Martin Underwood, Rachelle Buchbinder, Dan
  Cherkin, Nadine~E. Foster, Chris~G. Maher, Maurits {van Tulder}, Johannes~R.
  Anema, Roger Chou, Stephen~P. Cohen, Luc{\'i}ola {Menezes Costa}, Peter
  Croft, Manuela Ferreira, Paulo~H. Ferreira, Julie~M. Fritz, Douglas~P. Gross,
  Bart~W. Koes, Birgitta {\"O}berg, Wilco~C. Peul, Mark Schoene, Judith~A.
  Turner, and Anthony Woolf.
\newblock What low back pain is and why we need to pay attention.
\newblock \emph{The Lancet}, 391\penalty0 (10137):\penalty0 2356--2367, 2018.

\bibitem[Hastie et~al.(2009)Hastie, Tibshirani, and Friedman]{Hastie.2009}
Trevor Hastie, Robert Tibshirani, and J.~H. Friedman.
\newblock \emph{The Elements of Statistical Learning: Data Mining, Inference,
  and Prediction}.
\newblock Springer Series in Statistics. Springer, New York, 2 edition, 2009.
\newblock ISBN 978-0-387-84857-0.

\bibitem[Hatt and Feuerriegel(2022)]{Hatt}
Tobias Hatt and Stefan Feuerriegel.
\newblock Detecting user exits from online behavior: A duration-dependent
  latent state model.
\newblock \emph{arXiv:2208.03937}, 2022.

\bibitem[Helm et~al.(2015)Helm, Lavieri, {van Oyen}, Stein, and
  Musch]{Helm.2015}
Jonathan~E. Helm, Mariel~S. Lavieri, Mark~P. {van Oyen}, Joshua~D. Stein, and
  David~C. Musch.
\newblock Dynamic forecasting and control algorithms of glaucoma progression
  for clinician decision support.
\newblock \emph{Operations Research}, 63\penalty0 (5):\penalty0 979--999, 2015.

\bibitem[Henly(2017)]{Henly.2017}
Susan~J. Henly.
\newblock \emph{The Routledge international handbook of advanced quantitative
  methods in nursing research}.
\newblock Routledge, London, UK, 2017.
\newblock ISBN 9781138552852.

\bibitem[Hill et~al.(2020)Hill, Garvin, Chen, Cooper, Wathall, Bartlam,
  Saunders, Lewis, Protheroe, Chudyk, Birkinshaw, Dunn, Jowett, Oppong, Hay,
  {van der Windt}, Mallen, and Foster]{Hill.2020}
Jonathan Hill, Stefannie Garvin, Ying Chen, Vincent Cooper, Simon Wathall,
  Bernadette Bartlam, Benjamin Saunders, Martyn Lewis, Joanne Protheroe, Adrian
  Chudyk, Hollie Birkinshaw, Kate~M. Dunn, Sue Jowett, Raymond Oppong, Elaine
  Hay, Danielle {van der Windt}, Christian Mallen, and Nadine~E. Foster.
\newblock Computer-based stratified primary care for musculoskeletal
  consultations compared with usual care: Study protocol for the {STarT MSK}
  cluster randomized controlled trial.
\newblock \emph{JMIR Research Protocols}, 9\penalty0 (7):\penalty0 e17939,
  2020.

\bibitem[Hill et~al.(2008)Hill, Dunn, Lewis, Mullis, Main, Foster, and
  Hay]{Hill.2008}
Jonathan~C. Hill, Kate~M. Dunn, Martyn Lewis, Ricky Mullis, Chris~J. Main,
  Nadine~E. Foster, and Elaine~M. Hay.
\newblock A primary care back pain screening tool: Identifying patient
  subgroups for initial treatment.
\newblock \emph{Arthritis and Rheumatism}, 59\penalty0 (5):\penalty0 632--641,
  2008.

\bibitem[Horwitz et~al.(2018)Horwitz, Charlson, and Singer]{Horwitz.2018}
Ralph~I. Horwitz, Mary~E. Charlson, and Burton~H. Singer.
\newblock Medicine based evidence and personalized care of patients.
\newblock \emph{European Journal of Clinical Investigation}, 48\penalty0
  (7):\penalty0 e12945, 2018.

\bibitem[Jasra et~al.(2005)Jasra, Holmes, and Stephens]{Jasra.2005}
Ajay Jasra, Chris~C. Holmes, and David~A. Stephens.
\newblock Markov chain {M}onte {C}arlo methods and the label switching problem
  in {B}ayesian mixture modeling.
\newblock \emph{Statistical Science}, 20\penalty0 (1):\penalty0 50--67, 2005.

\bibitem[Jensen et~al.(2015)Jensen, Tom{\'e}-Pires, Sol{\'e}, Racine,
  Castarlenas, de~{La Vega}, and Mir{\'o}]{Jensen.2015}
Mark~P. Jensen, Catarina Tom{\'e}-Pires, Ester Sol{\'e}, M{\'e}lanie Racine,
  Elena Castarlenas, Roc{\'i}o de~{La Vega}, and Jordi Mir{\'o}.
\newblock Assessment of pain intensity in clinical trials: Individual ratings
  vs composite scores.
\newblock \emph{Pain Medicine}, 16\penalty0 (1):\penalty0 141--148, 2015.

\bibitem[Jin et~al.(2021)]{Jin.2021}
Lixia Jin et~al.
\newblock Predictive classification system for low back pain based on
  unsupervised clustering.
\newblock \emph{Global Spine Journal}, 13\penalty0 (3):\penalty0 1--6, 2021.

\bibitem[Joe(1997)]{Joe.1997}
Harry Joe.
\newblock \emph{Multivariate models and multivariate dependence concepts}.
\newblock {CRC Press}, Boca Raton, FL, 1997.
\newblock ISBN 0412073315.

\bibitem[Kongsted et~al.(2015)Kongsted, Kent, Hestbaek, and
  Vach]{Kongsted.2015}
Alice Kongsted, Peter Kent, Lise Hestbaek, and Werner Vach.
\newblock Patients with low back pain had distinct clinical course patterns
  that were~typically neither complete recovery nor constant pain: A latent
  class analysis of longitudinal data.
\newblock \emph{The Spine Journal}, 15\penalty0 (5):\penalty0 885--894, 2015.

\bibitem[Larsen(2017)]{Larsen.2017}
Pamala~D. Larsen.
\newblock \emph{Chronic Illness: Impact and Intervention}.
\newblock {Jones {\&} Bartlett Learning}, Burlington, 10 edition, 2017.
\newblock ISBN 978-1284128857.

\bibitem[Lee and van~der Schaar(2020)]{lee2020temporal}
Changhee Lee and Mihaela van~der Schaar.
\newblock Temporal phenotyping using deep predictive clustering of disease
  progression.
\newblock In \emph{International Conference on Machine Learning}, 2020.

\bibitem[Liu et~al.(2016)Liu, Mo, Song, and Wang]{liu2016longitudinal}
Yihao Liu, Shenjiang Mo, Yifan Song, and Mo~Wang.
\newblock Longitudinal analysis in occupational health psychology: A review and
  tutorial of three longitudinal modeling techniques.
\newblock \emph{Applied Psychology}, 65\penalty0 (2):\penalty0 379--411, 2016.

\bibitem[Liu et~al.(2023)Liu, Culpepper, and Chen]{liu2023identifiability}
Ying Liu, Steven~Andrew Culpepper, and Yuguo Chen.
\newblock Identifiability of hidden {M}arkov models for learning trajectories
  in cognitive diagnosis.
\newblock \emph{Psychometrika}, 88\penalty0 (2):\penalty0 361--386, 2023.

\bibitem[Maag et~al.(2021)Maag, Feuerriegel, Kraus, Saar-Tsechansky, and
  Z{\"u}ger]{Maag.2021}
Basil Maag, Stefan Feuerriegel, Mathias Kraus, Maytal Saar-Tsechansky, and
  Thomas Z{\"u}ger.
\newblock Modeling longitudinal dynamics of comorbidities.
\newblock In \emph{Conference on Health, Inference, and Learning (CHIL)}, 2021.

\bibitem[MacDonald and Zucchini(1997)]{MacDonald.1997}
Iain~L. MacDonald and Walter Zucchini.
\newblock \emph{Hidden Markov and other models for discrete-valued time
  series}.
\newblock {Chapman and Hall}, London, 1997.
\newblock ISBN 9780412558504.

\bibitem[Manduchi et~al.(2021)Manduchi, H{\"u}ser, Faltys, Vogt, R{\"a}tsch,
  and Fortuin]{Manduchi.2021}
Laura Manduchi, Matthias H{\"u}ser, Martin Faltys, Julia Vogt, Gunnar
  R{\"a}tsch, and Vincent Fortuin.
\newblock {T-DPSOM}: An interpretable clustering method for unsupervised
  learning of patient health states.
\newblock In \emph{Conference on Health, Inference, and Learning (CHIL)}, 2021.

\bibitem[Mueller-Peltzer et~al.(2020)Mueller-Peltzer, Feuerriegel, Nielsen,
  Kongsted, Vach, and Neumann]{Mueller.2020}
Michael Mueller-Peltzer, Stefan Feuerriegel, Anne~Molgaard Nielsen, Alice
  Kongsted, Werner Vach, and Dirk Neumann.
\newblock Longitudinal healthcare analytics for disease management: Empirical
  demonstration for low back pain.
\newblock \emph{Decision Support Systems}, 132:\penalty0 113271, 2020.

\bibitem[Murray et~al.(2020)Murray, Eisner, Nagin, and Ribeaud]{Murray2020}
Aja~L Murray, Manuel Eisner, Daniel Nagin, and Denis Ribeaud.
\newblock A multi-trajectory analysis of commonly co-occurring mental health
  issues across childhood and adolescence.
\newblock \emph{European Child \& Adolescent Psychiatry}, pages 1--15, 2020.

\bibitem[Murray et~al.(2022)Murray, Nagin, Obsuth, Ribeaud, and
  Eisner]{Murray2022}
Aja~Louise Murray, Daniel Nagin, Ingrid Obsuth, Denis Ribeaud, and Manuel
  Eisner.
\newblock Young adulthood outcomes of joint mental health trajectories: A
  group-based trajectory model analysis of a 13-year longitudinal cohort study.
\newblock \emph{Child Psychiatry \& Human Development}, 53\penalty0
  (5):\penalty0 1083--1096, 2022.

\bibitem[Nagin et~al.(2018)Nagin, Jones, Passos, and Tremblay]{Nagin2018}
Daniel~S Nagin, Bobby~L Jones, Valeria~Lima Passos, and Richard~E Tremblay.
\newblock Group-based multi-trajectory modeling.
\newblock \emph{Statistical Methods in Medical Research}, 27\penalty0
  (7):\penalty0 2015--2023, 2018.

\bibitem[Nagin et~al.(2024)Nagin, Jones, and Elmer]{Nagin2024}
Daniel~S Nagin, Bobby~L Jones, and Jonathan Elmer.
\newblock Recent advances in group-based trajectory modeling for clinical
  research.
\newblock \emph{Annual Review of Clinical Psychology}, 20, 2024.

\bibitem[Naumzik et~al.(2022)Naumzik, Feuerriegel, and Weinmann]{MKSC}
Christof Naumzik, Stefan Feuerriegel, and Markus Weinmann.
\newblock I will survive: Predicting business failures from customer ratings.
\newblock \emph{Marketing Science}, 41\penalty0 (1):\penalty0 188--207, 2022.

\bibitem[Naumzik et~al.(2023)Naumzik, Feuerriegel, and Nielsen]{EJOR}
Christof Naumzik, Stefan Feuerriegel, and Anne~Molgaard Nielsen.
\newblock Data-driven dynamic treatment planning for chronic diseases.
\newblock \emph{European Journal of Operational Research}, 305\penalty0
  (2):\penalty0 853--867, 2023.

\bibitem[{NHS Foundation Trust}(2015)]{RoyalMarsden.2015}
{NHS Foundation Trust}, editor.
\newblock \emph{Royal {M}arsden Manual of Clinical Nursing Procedures}.
\newblock {John Wiley {\&} Sons}, Somerset, UK, 9 edition, 2015.
\newblock ISBN 978-1-118-74592-2.

\bibitem[Nielsen et~al.(2016)Nielsen, Vach, Kent, Hestbaek, and
  Kongsted]{Nielsen.2016}
Anne~Molgaard Nielsen, W.~Vach, P.~Kent, L.~Hestbaek, and A.~Kongsted.
\newblock Using existing questionnaires in latent class analysis: Should we use
  summary scores or single items as input? {A} methodological study using a
  cohort of patients with low back pain.
\newblock \emph{Clinical Epidemiology}, 8:\penalty0 73--89, 2016.

\bibitem[Nielsen et~al.(2020)Nielsen, Binding, Ahlbrandt-Rains, Boeker,
  Feuerriegel, and Vach]{JCE}
Anne~Molgaard Nielsen, Adrian Binding, Casey Ahlbrandt-Rains, Martin Boeker,
  Stefan Feuerriegel, and Werner Vach.
\newblock Exploring conceptual preprocessing for developing prognostic models:
  A case study in low back pain patients.
\newblock \emph{Journal of Clinical Epidemiology}, 122:\penalty0 27--34, 2020.

\bibitem[Nikoloulopoulos(2013)]{Nikoloulopoulos.2013}
Aristidis~K. Nikoloulopoulos.
\newblock Copula-based models for multivariate discrete response data.
\newblock In \emph{Copulae in Mathematical and Quantitative Finance}, volume
  213 of \emph{Lecture Notes in Statistics}, pages 231--249. Springer,
  Heidelberg, 2013.
\newblock ISBN 978-3-642-35406-9.

\bibitem[Ozyurt et~al.(2021)Ozyurt, Kraus, Hatt, and Feuerriegel]{AttDMM}
Yilmazcan Ozyurt, Mathias Kraus, Tobias Hatt, and Stefan Feuerriegel.
\newblock {AttDMM}: An attentive deep {M}arkov model for risk scoring in
  intensive care units.
\newblock In \emph{ACM SIGKDD Conference on Knowledge Discovery \& Data Mining
  (KDD)}, 2021.

\bibitem[Panahiazar et~al.(2015)Panahiazar, Taslimitehrani, Pereira, and
  Pathak]{Panahiazar.2015}
Maryam Panahiazar, Vahid Taslimitehrani, Naveen~L. Pereira, and Jyotishman
  Pathak.
\newblock Using {EHR}s for heart failure therapy recommendation using
  multidimensional patient similarity analytics.
\newblock \emph{Studies in Health Technology and Informatics}, 210:\penalty0
  369--373, 2015.

\bibitem[Rabiner(1989)]{Rabiner.1989}
Lawrence~R. Rabiner.
\newblock A tutorial on hidden {M}arkov models and selected applications in
  speech recognition.
\newblock \emph{Proceedings of the IEEE}, 77\penalty0 (2):\penalty0 257--286,
  1989.

\bibitem[Rudin(2019)]{Rudin.2019}
Cynthia Rudin.
\newblock Stop explaining black box machine learning models for high stakes
  decisions and use interpretable models instead.
\newblock \emph{Nature Machine Intelligence}, 1\penalty0 (5):\penalty0
  206--215, 2019.

\bibitem[Schleidgen et~al.(2013)Schleidgen, Klingler, Bertram, Rogowski, and
  Marckmann]{Schleidgen.2013}
Sebastian Schleidgen, Corinna Klingler, Teresa Bertram, Wolf~H. Rogowski, and
  Georg Marckmann.
\newblock What is personalized medicine: Sharpening a vague term based on a
  systematic literature review.
\newblock \emph{BMC Medical Ethics}, 55\penalty0 (14):\penalty0 1--12, 2013.

\bibitem[Scott et~al.(2005)Scott, James, and Sugar]{Scott.2005}
Steven~L. Scott, Gareth~M. James, and Catherine~A. Sugar.
\newblock Hidden {M}arkov models for longitudinal comparisons.
\newblock \emph{Journal of the American Statistical Association}, 100\penalty0
  (470):\penalty0 359--369, 2005.

\bibitem[Shirley et~al.(2010)Shirley, Small, Lynch, Maisto, and
  Oslin]{Shirley.2010}
Kenneth~E. Shirley, Dylan~S. Small, Kevin~G. Lynch, Stephen~A. Maisto, and
  David~W. Oslin.
\newblock Hidden {M}arkov models for alcoholism treatment trial data.
\newblock \emph{The Annals of Applied Statistics}, 4\penalty0 (1):\penalty0
  366--395, 2010.

\bibitem[Silva et~al.(2022)Silva, Costa, Hancock, Palomo, Costa, and {Da
  Silva}]{Silva.2022}
Fernanda~G. Silva, Leonardo~Op Costa, Mark~J. Hancock, Gabriele~A. Palomo,
  Luc{\'i}ola~Cm Costa, and Tatiane {Da Silva}.
\newblock No prognostic model for people with recent-onset low back pain has
  yet been demonstrated to be suitable for use in clinical practice: A
  systematic review.
\newblock \emph{Journal of Physiotherapy}, 68\penalty0 (2):\penalty0 99--109,
  2022.

\bibitem[Strimbu and Tavel(2010)]{strimbu2010biomarkers}
Kyle Strimbu and Jorge~A Tavel.
\newblock What are biomarkers?
\newblock \emph{Current Opinion in HIV and AIDS}, 5\penalty0 (6):\penalty0 463,
  2010.

\bibitem[Vehtari et~al.(2017)Vehtari, Gelman, and Gabry]{Vehtari.2017}
Aki Vehtari, Andrew Gelman, and Jonah Gabry.
\newblock Practical {B}ayesian model evaluation using leave-one-out
  cross-validation and {WAIC}.
\newblock \emph{Statistics and Computing}, 27\penalty0 (5):\penalty0
  1413--1432, 2017.

\bibitem[Vos et~al.(2020)]{GBD.2019}
Theo Vos et~al.
\newblock Global burden of 369 diseases and injuries in 204 countries and
  territories, 1990--2019: a systematic analysis for the {Global Burden of
  Disease Study 2019}.
\newblock \emph{The Lancet}, 396\penalty0 (10258):\penalty0 1204--1222, 2020.

\bibitem[Zhang et~al.(2023)Zhang, Jin, Hartvigsen, Udler, and
  Ghassemi]{zhang2023pipeline}
Haoran Zhang, Qixuan Jin, Thomas Hartvigsen, Miriam Udler, and Marzyeh
  Ghassemi.
\newblock A pipeline for interpretable clinical subtyping with deep metric
  learning.
\newblock In \emph{ICML 3rd Workshop on Interpretable Machine Learning in
  Healthcare (IMLH)}, 2023.

\end{thebibliography}

\clearpage
\newpage
\appendix
\raggedbottom

\begin{table*}[h]
\centering
\LARGE Online Supplements
\end{table*}

\clearpage

\clearpage
\section{Estimation details}
\label{appendix:model_estimation}

\textbf{Bayesian estimation:} We estimate our \model using a so-called ``fully'' Bayesian approach \citep{Gelman.2014}. Specifically, we use Markov chain Monte Carlo in order to sample from the joint posterior distribution of the model parameters, and, for this, we later derive the likelihood $L$ for our \model. Because of this, our estimation approach is different from others: we do not need a maximum likelihood approach, an expectation-maximization algorithm, or a Metropolis-Hastings scheme; instead, for our ``fully'' Bayesian approach, we simply need to derive the likelihood $L$.

For an efficient calculation, we first implement the forward algorithm \citep{Rabiner.1989} in order to accelerate the computation. Second, the copula is directly integrated into the log-likelihood through a pair-wise construction scheme, which reduces the number of required evaluations. Third, dependence amongst risk factors can induce a strong posterior correlation, thereby rendering sampling from the posterior distribution ineffective. We address this by centering the risk factors and then transforming them into a set of linearly uncorrelated variables by applying a QR decomposition. Mathematically, the matrix of risk factors $X = QR$ is decomposed into an orthogonal matrix $Q$ and an upper triangular matrix $R$, where the cluster weights are parameterized through $\tilde{\beta}^k = R\beta^k$. Fourth, we address potential label switching in latent models by following the suggestions from \citet{Jasra.2005}.

\textbf{Priors:} The priors for all model parameters are set to weakly informative priors. For the coefficients inside the cluster membership, we choose zero mean normal priors with standard deviation equal to $5$ for the intercepts $\alpha^k$ and equal to $1$ for $\tilde{\beta}^k$. To ensure identifiability of the model, we set the free parameters for one subgroup (\ie, $\alpha^1$ and $\beta^1$) to zero. We choose a symmetric Dirichlet distribution as prior for the initial state distribution $\pi^k$ and the transition probabilities $\Phi^k$. For the initial state distribution, we set all parameters to $1$, except for state~1 (``severe'') where the parameter was set to $S$, thereby incorporating the fact that all patients in our study have just consulted a medical professional. For the transition probabilities, we set the parameters to $1$, except for the diagonal elements where we again set it to $S$ as well, in order to penalize frequent switching between the hidden states. Consistent with the literature, we model the Likert-based values for our symptoms as a truncated Poisson distribution and their marginal distributions $F_P$ and $F_D$ are parameterized by their corresponding means $\lambda_P^k=\left(\lambda_{P1}^k,\ldots,\lambda_{P\mathcal{S}}^k\right)$ and $\lambda_D^k=\left(\lambda_{D1}^k,\ldots,\lambda_{D\mathcal{S}}^k\right)$, respectively. For the pain and disability parameters $\lambda_P^k$ and $\lambda_D^k$, we choose a truncated normal distribution with a standard deviation of $5$. Finally, for the copula parameters $\rho^k$, we place a truncated normal distribution with standard deviation of $5$ on $\tilde{\rho}^k$ with $\rho^k = 1 + \tilde{\rho}^k$.

\textbf{Sampling:} Our implementation further draws upon recent advances in Bayesian estimations \citep{Gelman.2014}, namely, the Hamiltonian Monte Carlo algorithm from the software \textquote{Stan} with additional optimization via the No-U-Turn~(Nuts) sampler. This approach differs from other estimation techniques, specifically the Metropolis-Hastings algorithm or maximum likelihood estimation. Different from them, our estimation approach based on the Hamiltonian Monte Carlo sampler is considerably more efficient \citep{Gelman.2014}. Typically, it requires fewer chains/iterations by several orders of magnitude \citep{Gelman.2014}. We first ran four chains, each with 2,000 iterations of which we excluded the first 1,000 samples as part of a warm-up. We then checked that the average likelihoods in each chain were sufficiently close to each other. We finally use a single chain with 5,000 samples to generate the results for all subsequent evaluations. Here, we again exclude the initial 1,000 samples as part of a warm-up. We initialize all parameters with maximum likelihood estimates in order to speed up sampling. 

\textbf{Model diagnostics:} We follow common practice in Bayesian modeling \citep{Gelman.2014} by performing the following model diagnostics. This is to ensure convergence of the MCMC algorithm and thus precise estimates. (1)~We inspected the effective sample size $n_\text{eff}$, indicating that the number of MCMC samples is sufficient. (2)~We calculated the Gelman-Rubin convergence diagnostic $\hat{R}$ of all model parameters. The $\hat{R}$ is below the critical threshold of 1.02, suggesting convergence of the MCMC chains. (3)~We manually inspected traceplots. The trace plots suggest that the chains have mixed well. (4)~We also validated our model design by testing whether we can retrieve the parameters from simulated data. All checks had the desired outcomes.

\textbf{Computational performance:} We conducted our experiments using standard office hardware (Intel i7-8550U 8th generation CPU with 16 GB RAM) to mimic typical computational infrastructure in clinical settings. The runtime is $\sim$24\,h but we note that such training is only done once while the inference time for assigning incoming patients to subgroups is $<$1\,s. We further emphasize the following. First, discussions with the medical professionals from our author team suggest that the runtime is acceptable. The reason is that such analysis is done only once and for comparatively small cohorts where the patient characteristics are not yet properly understood but where new disease markers should be identified. Second, we opted for a fully Bayesian approach based on MCMC sampling to obtain posterior estimates for all coefficients. Needless to say, variational inference can effectively reduce the runtime further if desired.  

\clearpage
\section{Dataset details}
\label{appendix:dataset}

This work builds upon an extensive, longitudinal study of 847 patients with non-specific low back pain \citep{Kongsted.2015}. Initially, 928 patients participated in the study, but we excluded 81 of them from our sample based on feedback from the clinical practitioners, as the patients failed to respond to weekly monitoring with sufficient quality. We conducted additional robustness checks by re-estimating our model with two sub-samples, namely those patients for whom all weekly measurements were obtained and only patients with missing values. However, we find no notable differences across both sub-samples. The actual design of the study was developed and undertaken by experts from clinical back pain research who are part of our interdisciplinary author team. Specifically, patients aged between 18 and 65 years were approached to participate in the cohort study. They had to present to a chiropractor with low back pain as their primary reason for care-seeking at the time of enrollment. 

The duration of the clinical study for data collection was 52 weeks, as this exceeds the usual length of low back pain episodes by several orders \citep{Kongsted.2015}. In our study, \SI{62.54}{\percent} of all low back pain episodes last up to 2~weeks, \SI{13.69}{\percent} between 2--4 weeks and \SI{10.68}{\percent} between 1--3 months. Hence, a length of 52 weeks for our longitudinal study is sufficient to capture multiple episodes with (severe) low back pain.

Data collection was three-fold:
\begin{enumerate}
\item A comprehensive upfront survey was used to collect baseline variables with potential risk factors $x_i$. Baseline variables link to the risk of onset of low back pain and may thus partially explain the heterogeneity across patients \citep{JCE}. We use the term ``risk factor'' throughout our paper as this is consistent with other works in health management and thereby highlights the generalizability of our model. However,  the term ``prognostic factor'' may be preferred in some disciplines for factors used to predict prognosis in a present health condition, reserving ``risk'' to the risk of acquiring a disease. The baseline variables include information on socioeconomic characteristics, such as gender and occupation, as well as pre-existing clinical conditions (see Table~\ref{tbl:list_risk_factors}). The actual choice was made by the clinical researchers from our author team according to best practice in clinical research. We followed common conventions in the design of medical studies where such an extensive catalog of baseline variables is only collected once.
\item The progression of low back pain was monitored through weekly follow-up questions. These capture the symptoms $y_{it}$. Here, we collected the typical pain intensity according to the NRS scale (0--10), which is the standard for a valid and reliable assessment of pain in healthcare research \citep{Breivik.2008}, and, further, the number of days with activity limitation during that week. 
\item A final, follow-up questionnaire was used to further collect health outcomes after 12 months for pain (using the NRS scale) and disability (using the Roland-Morris Disability Questionnaire on a 0--100 scale).   
\end{enumerate}

\begin{table*}[h]
\centering

\footnotesize
\begin{tabular}{ll}
\toprule
Risk factor & Range/levels \\
\midrule
Age (in years) & 18--65 \\
Body mass index (BMI) & 18--59 \\
Current pain level & 0--10 \\
General health & 0--100 \\
Height (in cm) & 153--201 \\
Back pain dominating & Yes/No \\
Days with low back pain last year & Less than 30/More than 30 \\
Duration of pain episode & 0--2 weeks/2--4 weeks/1--3 months/More than 3 months\\
Educational level & No vocational education/Vocational education\\
&Short/Medium/Long higher education\\
Gender & Female/male \\
Other chronic disease & Yes/No \\
Pain distribution & Back pain only/Leg pain only \\
	&Back pain and pain in one leg/\\
	&Back pain and pain in both legs\\
Physical work load & Sitting/Sitting and walking/Light/Heavy physical work\\
Previous low back pain episodes & None/1--3/More than 3 \\
Severity of leg pain & No pain/Mild pain/Moderate-severe pain\\
Smoking status & Smoker/Ex-smoker/Never smoked \\
Work situation & Unemployed/Student/Self-employed/ \\
	&Part-time/Full-time/Retired/Other\\
\bottomrule
\end{tabular}
\caption{List of risk factors used in this study. }
\label{tbl:list_risk_factors}
\end{table*}

The dataset comprises 847 patients with a mean age of \num{43.2} years and out of which 389 (\SI{45.93}{\percent}) are women. The mean pain value is 1.45 (SD of \num{2.21}), while the mean disability amounts to \num{0.79} days with activity limitations per week (SD of \num{1.85}). Naturally, both dimensions are correlated, as shown by a correlation coefficient of \num{0.69} with a statistically significant $p$-value of below \num{0.0001}.

We randomly split the dataset into an almost equally-sized training set (425 subjects) and test set (remaining 422 subjects). The former is used to train the \model, while the latter is used for out-of-sample performance evaluations.

\clearpage
\section{Estimation of our \model: Model selection to determine the number of subgroups}
\label{appendix:estimation_model}

We now identify the number of subgroups for our \model. We estimated \models with different numbers of subgroups $K$ and then compared the model fit. The \models are fit to the dataset from our clinical study above, namely, static variables and health trajectories over 52 weeks from 847 patients. As is common in Bayesian modeling, we report both the in-sample and out-of-sample log pointwise predictive density~(lpd) \citep{Vehtari.2017}. We further follow best practice in Bayesian modeling according to which the final model selection should be based on the out-of-sample lpd \citep{Gelman.2014}. The reason is that this metric captures how well the model generalizes to unseen patients and is thus a measure of model accuracy \citep{Vehtari.2017}. Previous research on Bayesian modeling \citep[cf.][]{MKSC,Hatt} has frequently utilized the deviance information criterion~(DIC) or the Akaike information criterion~(AIC). However, both rely only on single point estimates, which makes them often meaningless in the context of mixture and hierarchical models \citep{Gelman.2014}. We circumvent this by using the out-of-sample lpd from \citet{Vehtari.2017}, since these explicitly incorporate the whole posterior distribution.

The results are presented in Table~\ref{tbl:mhmm_cluster_results}. It compares our \model with multivariate health trajectories based on our copula approach. Specifically, the lpd assesses predictive accuracy in a Bayesian framework and, therefore, the overall fit of a model \citep{Gelman.2014,Vehtari.2017}. A lower lpd (deviance scale) indicates a better fit. For our \model, the lowest in-sample lpd is found for $K = 8$ subgroups suggesting that this gives the best in-sample fit for our data. Further, we look at the out-of-sample lpd for model selection. The lowest out-of-sample lpd is achieved for $K = 8$ subgroups. Hence, our \model with $K = 8$ performs best, so that this choice should be used as a result of the model selection. Therefore, all subsequent interpretations are based on $K = 8$ subgroups. 

\begin{table}[h]
  \centering
	{\
	 \footnotesize
		\begin{tabular}{c SS}
			\toprule
			\#Subgroups $K$ & \multicolumn{2}{c}{Lpd}\\
			\cmidrule(lr){2-3} 
			 & \multicolumn{1}{c}{In-sample} & \multicolumn{1}{c}{Out-of-sample} \\
			\midrule
			 \csname @@input\endcsname number_of_clusters.tex
			\bottomrule	
			\end{tabular}
			}
	\caption{Comparison of our \model subgrouping across a varying number of subgroups $K$ in order to perform model selection. The final selection is based on the out-of-sample log pointwise predictive density \citep{Gelman.2014,Vehtari.2017}, which measures the ability how well our model generalizes to unseen patients and is thus a measure of model accuracy. Hence, $K = 8$ subgroups are preferred.}
	\label{tbl:mhmm_cluster_results}
\end{table}

\clearpage
\section{Interpretation of subgroups}
\label{appendix:extended_results}

We now discuss how the different subgroups obtained by our model lend to potentially clinically-relevant interpretation. For this, Table~\ref{tbl:interpretation_clusters} presents a descriptive overview of the identified subgroups, where, along for each subgroup, we further provide names and descriptions that were proposed by the healthcare experts in the field of low back pain (see Supplement~\ref{appendix:conceptual_labeling} for our naming procedure). The table also includes statistics related to the health trajectory and the overall health outcomes after 12 months. The latter were collected from the patients during a final questionnaire subsequent to completing the study after $12$ months. Evidently, we observe considerable differences regarding the prevalence of subgroups, as well as their expected chance of recovery. For instance, the average pain level in the final questionnaire of subgroup~1 amounts to \num{5.33} on the NRS scale (0--10), while it amounts to only \num{1.15} for subgroup~3.

\thispagestyle{empty}
\begin{table}[h!]
\vspace{-1.2cm}
\centering
\rotatebox{90}
{

\tiny
\begin{tabular}{c p{1.5cm} p{2.5cm} S SSSSSSSS SS}
\toprule 
\multicolumn{4}{c}{Subgroup}& \multicolumn{4}{c}{Risk factors} & \multicolumn{4}{c}{Longitudinal data} & \multicolumn{2}{c}{Outcomes} \\
\cmidrule(lr){1-4}\cmidrule(lr){5-8}\cmidrule(lr){9-12}\cmidrule(lr){13-14}
No. & Name & Description & \multicolumn{1}{c}{\#Patients} & \multicolumn{1}{c}{Male} & \multicolumn{1}{c}{Age} & \multicolumn{1}{c}{BMI} & \multicolumn{1}{c}{Mostly} & \multicolumn{2}{c}{Pain} & \multicolumn{2}{c}{Activity limitation} & \multicolumn{1}{c}{Pain} & \multicolumn{1}{c}{Disability} \\
\cmidrule(lr){9-10}\cmidrule(lr){11-12}
& & & & \multicolumn{1}{c}{(in \%)} & & & \multicolumn{1}{c}{physical} & \multicolumn{1}{c}{Mean} & \multicolumn{1}{c}{SD} & \multicolumn{1}{c}{Mean} & \multicolumn{1}{c}{SD} & \multicolumn{1}{c}{after 12} & \multicolumn{1}{c}{after 12} \\
& & & & & & & \multicolumn{1}{c}{work (in \%)} & & & & &  \multicolumn{1}{c}{months} & \multicolumn{1}{c}{months} \\
\midrule
\csname @@input\endcsname cluster_description.tex
\bottomrule
\multicolumn{14}{l}{BMI: body mass index, SD: Standard deviation}
\end{tabular}
}
\caption{Results of proposed patient subgrouping. This table reports additional descriptive statistics across common risk factors (based on the initial survey), the health trajectory (based on the weekly monitoring), and the health outcomes (based on the follow-up after 12 months) in each cluster. Statistics are reported as means. The description of the clusters was proposed by experts in the healthcare domain with specialization in low back pain. }
\label{tbl:interpretation_clusters}
\end{table}

While the previous descriptive statistics reported on the observed levels of pain and activity limitation, we now additionally recover the latent states in the health trajectory and discuss them in light of the Corbin-Strauss framework. For this, we calculated the likeliest latent state sequence via the Viterbi algorithm \citep[see, \eg,][]{Rabiner.1989} for each patient based on the respective subgroup-specific model parameters. Overall, the different states vary in their typical emissions. State~1 has the highest average pain intensity and activity limitation, followed by state~2 and state~3. Hence, the states may be interpreted through the lens of the Corbin-Strauss framework \citep{Corbin.1988,Corbin.1991,Corbin.1998}, and we refer to them as ``severe'', ``moderate'', and ``mild'' phases. In the following, we add the previous terms in quotation marks next to the state numbers to facilitate readability.  

Fig.~\ref{fig:patient_cluster} presents the relative frequency of the three latent states. These shed further light on the disease progressions among subgroups. As we can see, there is considerable heterogeneity across the different subgroups. For instance, fewer than \SI{50}{\percent} of the subjects in subgroups 2 and 7 are in state~3 (``mild'') while many are still in state~1 (``severe'') by the end of the observation period suggesting that many patients experience a relapse. Still, there are nuanced differences between subgroups 2 and 7 such as the latter having, on average, higher pain ratings and more pronounced activity limitations.  In contrast, subgroups 3 and 8 are characterized by a fairly rapid transition towards state~3 (``mild''): both entail the highest proportion of patients in state~3 (``mild''), with around \SI{80}{\percent} each.

As an additional illustration, we point towards the difference between subgroups 3 and 5. In fact, both subgroups feature similar progression behavior during the first five weeks of the monitoring, as the share of patients in state~3 (``mild'') rises sharply to around \SI{50}{\percent}. Beyond that, the trajectories differ: subgroup~3 recovers further until almost \SI{80}{\percent} of the subjects are in state~3 (``mild''). However, for subgroup~5, the proportion of state~3 (``mild'') evolves differently. It also continues to rise to about \SI{75}{\percent} after \num{26} weeks. Afterward, it decreases as more patients show a relapse and transition to state~2 (``moderate''). By the end of the study, the share of patients in state~3 (``mild'') in this subgroup is only around \SI{70}{\percent}. This is an interesting observation, since our subgrouping can accurately assign patients to either one of the two clusters and thereby offers prognostic power on whether patients experience a temporary recovery and thus the future disease dynamics. Conversely, the same would be difficult for healthcare practitioners using existing guidelines as the heterogeneity in the progression of low back pain is poorly understood.

Finally, the clinical researchers provided examples of subgroup-specific treatment plans. For instance, subgroup~3 entails mostly stable phases, a low average pain level with little volatility, and a positive prognosis. This subgroup thus would require only minimal treatment in the form of self-management advice. In contrast, subgroup~7 suffers from a high probability of moving to state 1 (``severe''), combined with intense pain and a negative prognosis. Hence, it mostly includes patients at risk for chronic problems and thus requires more extensive treatment or support.

\clearpage

\section{Empirical results for determining the optimal number of latent states ($\mathcal{S} = 3$)}
\label{appendix:robustness_states}

Here, we justify our choice of using $\mathcal{S} = 3$ latent states. Recall that prior theory from the so-called trajectory framework \citep{Corbin.1988,Corbin.1991,Corbin.1998} stipulates the presence of three latent states, which we now validate empirically. For this purpose, we fit models with a varying number of latent states, $\mathcal{S}$, and then compare the model fit. 

We follow state-of-the-art recommendations on evaluating Bayesian models and model selection \citep{Vehtari.2017}. We report both the in-sample and out-of-sample log pointwise predictive density~(lpd) \citep{Vehtari.2017}. We further follow best practice in Bayesian modeling according to which the final model selection should be based on the out-of-sample lpd \citep{Gelman.2014}. The reason is that this metric captures how well the model generalizes to unseen patients and is thus a measure of model accuracy \citep{Vehtari.2017}. Specifically, the lpd assesses predictive accuracy in a Bayesian framework, and, therefore, the overall fit of a model \citep{Gelman.2014,Vehtari.2017}. A lower lpd (deviance scale) indicates a better fit.

The results are presented in Table~\ref{tbl:mhmm_cluster_states}. The dataset is the same as that of the main paper, namely static variables and health trajectories over 52 weeks from 847 patients. We observe that including more than a single latent state to the model drastically improves the model fit. The model with $\mathcal{S}=1$ states corresponds to the autoregressive models for describing low back pain trajectories in \citet{Mueller.2020}. The model with $\mathcal{S}=1$ has the largest in-sample and out-of-sample lpd, and thus the worst fit out of the models under comparison. In comparison, the lowest out-of-sample lpd is observed for the model with $\mathcal{S}=3$, implying that this model performs best in describing our data. This finding is further corroborated by that the lowest in-sample lpd is also observed by the model with $\mathcal{S}=3$. Overall, three latent states yield the best fit for both in-sample and out-of-sample lpd. This finding confirms our initial hypothesis, and, accordingly, three states are used in all analyses.

\begin{table}[H]
\centering
\footnotesize
\begin{tabular}{l rr}
\toprule
$\mathcal S$ & \multicolumn{2}{c}{lpd}\\
\cmidrule(lr){2-3} 
 & {In-sample} & {Out-of-sample} \\
\midrule
\csname @@input\endcsname number_of_states.tex
\bottomrule	
\end{tabular}
\caption{Model fit across a different number of states. The choice of three latent states yields the best fit to the data across all of the considered metrics.}
\label{tbl:mhmm_cluster_states}
\end{table}

\clearpage
\section{Choice of survival Gumbel copula}
\label{appendix:robustness_copula}

We now offer an in-depth explanation of why our survival Gumbel copula is preferred using both theoretical and empirical arguments.

\textbf{Theoretical justification:} Health management commonly monitors the progression of diseases along multiple symptoms \citep[\eg,][]{Jensen.2015,liu2016longitudinal}; however, symptoms are usually not unrelated. Rather, they co-occur in a specific manner: (1)~either \emph{all} (or almost all) symptoms are absent when the patient has recovered, or (2) the condition is indicated by \emph{some} -- but not necessarily all -- symptoms due to differences in how patients respond to a disease. For instance, patients with stable low back pain experience an absence of both pain and activity limitation, whereas acute low back pain is usually characterized by severe pain \emph{or} severe activity limitation, though rarely both \citep{EJOR}. In other words, the absence of one characteristic makes it more likely that other characteristics will also be absent. Altogether, this results in a lower tail dependence among health measurements that must be modeled accordingly.

Our copula $C_s$ should accommodate tail dependence, so that an absence of symptoms appears jointly. In order to model this behavior, we draw upon a survival Gumbel copula. For $u,v\in[0,1]$, it is given by
{\small\begin{equation}
C_{\rho^k}(u,v) = u + v -1  
+ \exp\left[-\left( (-\log{1-u})^{\rho^k} + (-\log{1-v})^{\rho^k} \right)^{\frac{1}{\rho^k}}\right]
\end{equation}}
where the parameter $\rho^k \geq 1$ controls the strength of the tail dependence. It can be shown that the survival Gumbel copula has positive lower tail dependence for $\rho^k>1$ and zero upper tail dependence for all $\rho^k$ \citep[\eg,][]{Joe.1997}. For the special case of $\rho^k = 1$, the survival Gumbel copula reduces to independent observations.

\textbf{Empirical analysis:} In our analyses, we tested a variety of copulas. Specifically, we used the VineCopula package for R to compare a range of alternative copulas numerically, namely, tawn type II, BB7, Fran, and Joe, but found that the survival Gumbel copula gave the best empirical fit. Hence, we ultimately decided upon a survival Gumbel copula. We also re-estimated our model with other copulas such as the Ali-Mikhail-Haq copula in order to test for a symmetric dependence structure. It is given by $C_s(u, v) = u\, v\, (1 - \rho^k (1 - u) (1 - v))^{-1}$ with an additional parameter $\rho^k\in [-1, 1)$. Here the symmetric dependence structure follows from its multiplicative form. However, our numerical experiments confirmed that the Gumbel copula is superior.

\clearpage
\section{Sensitivity to time-varying risk factors}
\label{appendix:robustness_time-varying_factors}

Some risk factors may be time-varying, and we carefully tested that our results are not sensitive to them. Specifically, some of the baseline variables describe the prior history of patients, while others could theoretically be subject to variation over time; e.g. the work situation could change. We followed common conventions in the design of medical studies where such an extensive catalog of baseline variables is only collected once and, as a limitation, is not updated at regular time intervals. Nevertheless, we conducted a robustness check whereby we re-estimated our model but excluded all risk factors that could theoretically be subject to change (e.g. BMI or work situation); yet we obtained similar outcomes. If available in practice, further dynamic variables can be directly fed into our model, since, in contrast to other models, it supports multivariate observations with dependence structure as one of its strengths.

\clearpage
\section{Cluster validity indices} 
\label{appendix:cluster_validity_indices}

We compare the clustering performance in terms of so-called cluster validity indices~\citep{Arbelaitz.2013}. Cluster validity indices have been specifically developed to quantify the similarity of samples within a cluster (\ie, cluster cohesion) and their difference to other clusters (\ie, cluster separation). Previous research in the field of clustering has devised a variety of cluster validity indices as no metric is universally applicable. For this reason, we draw upon the following, conventional metrics that performed best in earlier research \citep{Arbelaitz.2013}, namely, the Calinski-Harabasz, Silhoutte, and Davies-Bouldin indices. These are introduced in the following (the arrow is used to indicate whether larger or smaller values are preferred): 
\begin{enumerate}
\item \textbf{Calinski-Harabasz}~(CH$\uparrow$): The clustering $\mathcal{C}=\{c_1,\ldots,c_K\}$ provides a partition of the set of patients, \ie, $\bigcup_{j=1}^Kc_j=\{1,\ldots,N\}$ and $c_i\cap c_j = \varnothing$ for $i\neq j$. For a cluster $c_k$, we define its {centroid} as the mean trajectory over all patients in the cluster, \ie, $\bar{c_k}=\abs{c_k}^{-1}\sum_{i\in c_k}y_i$. Similarly, the {global centroid} $\bar{y}$ is defined as the mean trajectory over all patients in the sample. Cluster cohesion is then assessed based on the distances of the trajectories in a cluster to its centroid, given by $\norm{y_i-\bar{c_k}}$. Cluster separation is measured via the distances of the cluster centroids to the global centroid, calculated as $\norm{\bar{y}-\bar{c_k}}$. This yields
\begin{equation}
\mathrm{CH\uparrow}(\mathcal C) = \frac{N-K}{K-1}\frac{\sum_{k=1}^{K}\abs{c_k}\norm{\bar{y}-\bar{c_k}}}{\sum_{k=1}^K\abs{c_k}S(c_k)}
\end{equation}
with $S(c_k) = \abs{c_k}^{-1}\sum_{i\in c_k}\norm{y_i-\bar{c_k}}$.
\item \textbf{Silhouette}~(Sil$\uparrow$): Cluster cohesion of the clustering $\mathcal C$ is measured based on the average distance between trajectories in the same cluster. Cluster separation is calculated as the minimum distance between two trajectories from different clusters. This yields
\begin{equation}
\mathrm{Sil\uparrow}(\mathcal C) = \frac{1}{N}\sum_{k=1}^K\sum_{i\in c_k}\frac{b(y_i,c_k)- a(y_i,c_k)}{\max\{a(y_i,c_k),\,b(y_i,c_k)\}}
\end{equation}
with $a(y_i,c_k) = \abs{c_k}^{-1}\sum_{j\in c_k}\norm{y_i-y_j}$ and $b(y_i,c_k)= \min_{c_l\in\mathcal C\setminus c_k}\{\abs{c_l}^{-1}\sum_{j\in c_l}\norm{y_i-y_j}\}$.
\item \textbf{Davies-Bouldin}~(DB*$\downarrow$): This index measures the cluster cohesion based on the distance of the trajectories in a cluster from its centroid and the cluster separation based on the distance between centroids. It is defined as
\begin{equation}
\mathrm{DB^\ast\downarrow}(\mathcal{C}) = \frac{1}{K}\sum_{k=1}^K\frac{\max_{l\neq k}\{S(c_k)+S(c_l)\}}{\min_{l\neq k}\{\norm{\bar{c_l}-\bar{c_k}}\}}.
\end{equation}
\end{enumerate}

\clearpage
\section{Conceptual naming}
\label{appendix:conceptual_labeling}

We assigned names and descriptions to each cluster by following a principled, three-step procedure. After computing the different subgroups, we prepared the following information for each subgroup: the average proportion of men/women, the average age, the BMI, and key statistics about physical work (as some subtypes of low back pain may be related to intense physical labor). We further computed average statistics for the patient trajectories with regard to pain intensity and activity limitation (including uncertainty estimates). We also retrieved information on the actual health outcomes from the follow-up procedure, that is, 12 months after the beginning of our clinical study. Information on health outcomes is not used during clustering but allows us to assess whether patients belonging to the same subgroup are sufficiently similar in terms of patient outcome. Then, the combined information was shown to one co-author (AK) and an additional clinical researcher from low back pain. The first one generated a summary description that should specifically point to the clinical relevance and, afterward, the description was checked by the second person. Thereby, we ensure that the descriptions provide accurate and meaningful characterizations of the underlying disease dynamics.

\clearpage
\section{Clinical considerations}
\label{appendix:online_offline_assignment}

\textbf{Benefits of probabilistic models:} There are two important benefits for probabilistic modeling. 

First, probabilistic models can identify the optimal number of clusters through an information criterion. Hence, the optimal number of clusters can be learned from data. In contrast, non-probabilistic methods (e.g., $k$-means) must rely on heuristics such as the elbow curve. Hence, this is a key advantage for using probabilistic models in practice.

Second, a probabilistic model allows practitioners to assess the confidence with which a patient is matched to a subgroup. This is different from non-probabilistic modeling, as can be seen in the following comparison:
\begin{itemize}
\item Non-probabilistic models (e.g., $k$-means or many of the other baselines) simply assign each patient to one subgroup. Hence, practitioners do not yield statistical estimates of how well such patients can be assigned to a subgroup. You may argue that we could simply measure the distance to the different centroids of the clusters but, in medicine, some scales are not directly comparable (e.g., is a $+10$ in age the same as $+10$ in BMI?). So, practitioners cannot yield insights into how well patients are matched to subgroups. 
\item Probalistic models estimate a probability with which patients are matched to a subgroup (e.g., a patient may be assigned to a subgroup with 98\% probability). Hence, practitioners can understand the confidence of the model to ensure that reliable assignments are made.
\end{itemize}
Later, this section demonstrates the value of having a probabilistic modeling approach. Therein, we plot the probability of the assigned subgroup (i.e., the highest probability out of the high probabilities for each cluster). For a non-informative clustering (``random setting''), each of the 8 subgroups would be equally likely and thus the maximum probability for one of the 8 clusters is 1/8 = 12.5\%. In contrast, we can see that our probabilistic model can accurately assign patients to one cluster as most probabilities are way above 50\%. Hence, all patients can be reliably and with great confidence matched to one subgroup.  

\begin{figure}[h]
	\centering
	\includegraphics[width=0.9\linewidth]{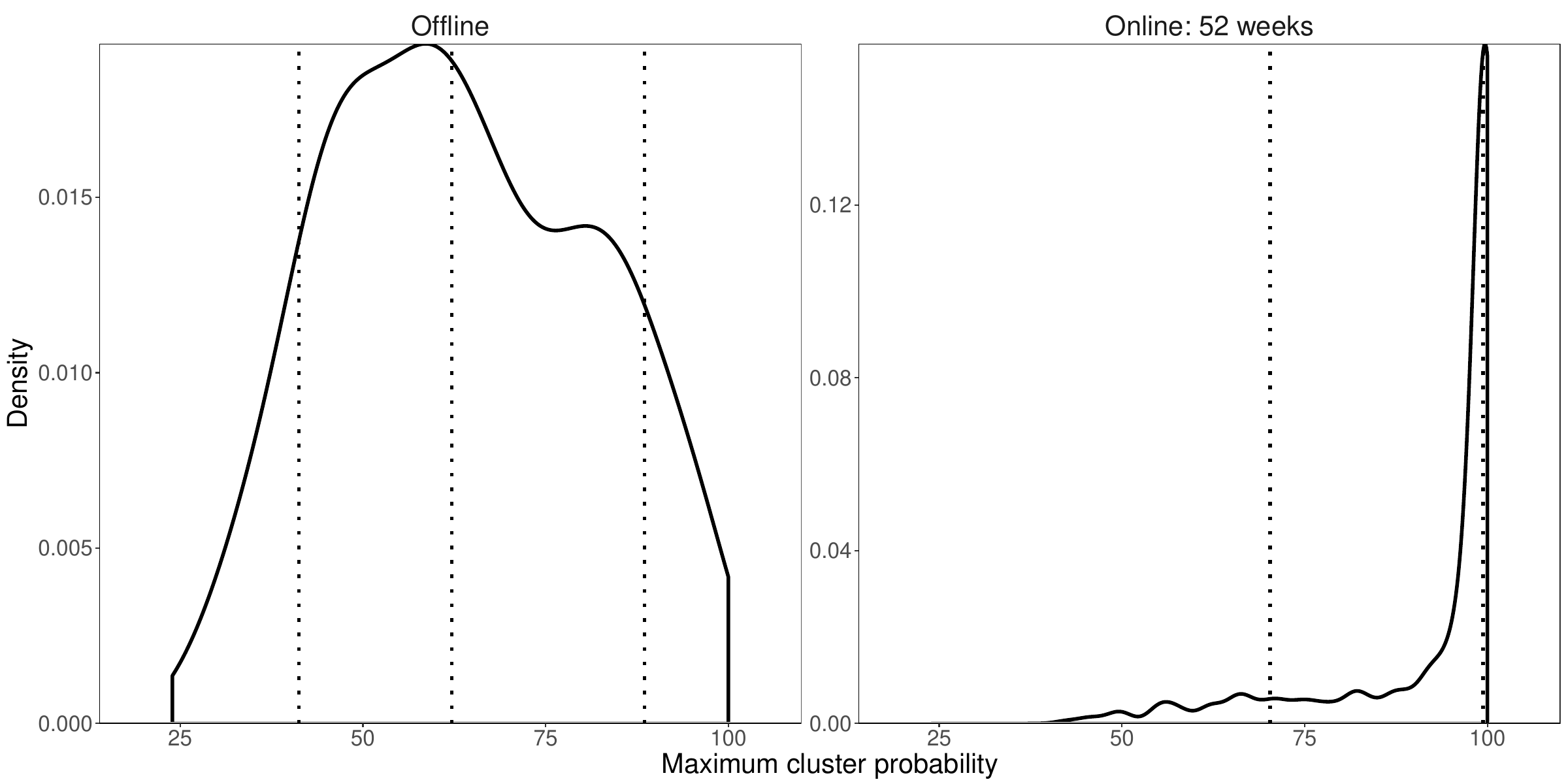}
	\caption{Kernel density plot of highest subgroup probabilities for each patient in the test set. The plot is for both offline~(left) and online~(right) assignment. The dotted lines mark the \SI{10}{\percent}, \SI{50}{\percent} and \SI{90}{\percent} quantiles. Hence, patients can almost always be matched to a single subgroup. }
	\label{fig:cluster_weights}
\end{figure}

\textbf{Confidence across online/offline assignment:} Our proposed approach allows for both offline and online subgroup assignment of patients. While the former approach relies only on the risk factors associated with each patient, the online approach incorporates additionally the trajectory data. Both approaches are inherently probabilistic in the sense that there is an estimated probability that a patient belongs to a specific subgroup. Ideally, one would like that, for each patient, there is only one subgroup with a large probability (and not that multiple subgroups are equally likely). This would imply that there is a clear matching towards one of the subgroups. Furthermore, the distribution of different probabilities should become less dispersed and thus more ``precise'', as a longer trajectory is observed. Therefore, we now compare the distribution of the maximum subgroup probabilities across the patients in the test set in Fig.~\ref{fig:cluster_weights}. Obviously, the offline subgroup probabilities exceed \SI{40}{\percent} for \SI{90}{\percent} of the patients, with a median probability of more than \SI{60}{\percent}. This indicates that the offline approach allows for a unique assignment of patients to subgroups. Moreover, the subgroup probabilities are even more ``sharp'' in the case of the online approach, where \SI{90}{\percent} of the patients correspond to a probability of above \SI{70}{\percent}, with a median probability of almost \SI{100}{\percent}.

\begin{figure}[h]
	\centering
	\includegraphics[width=0.95\linewidth]{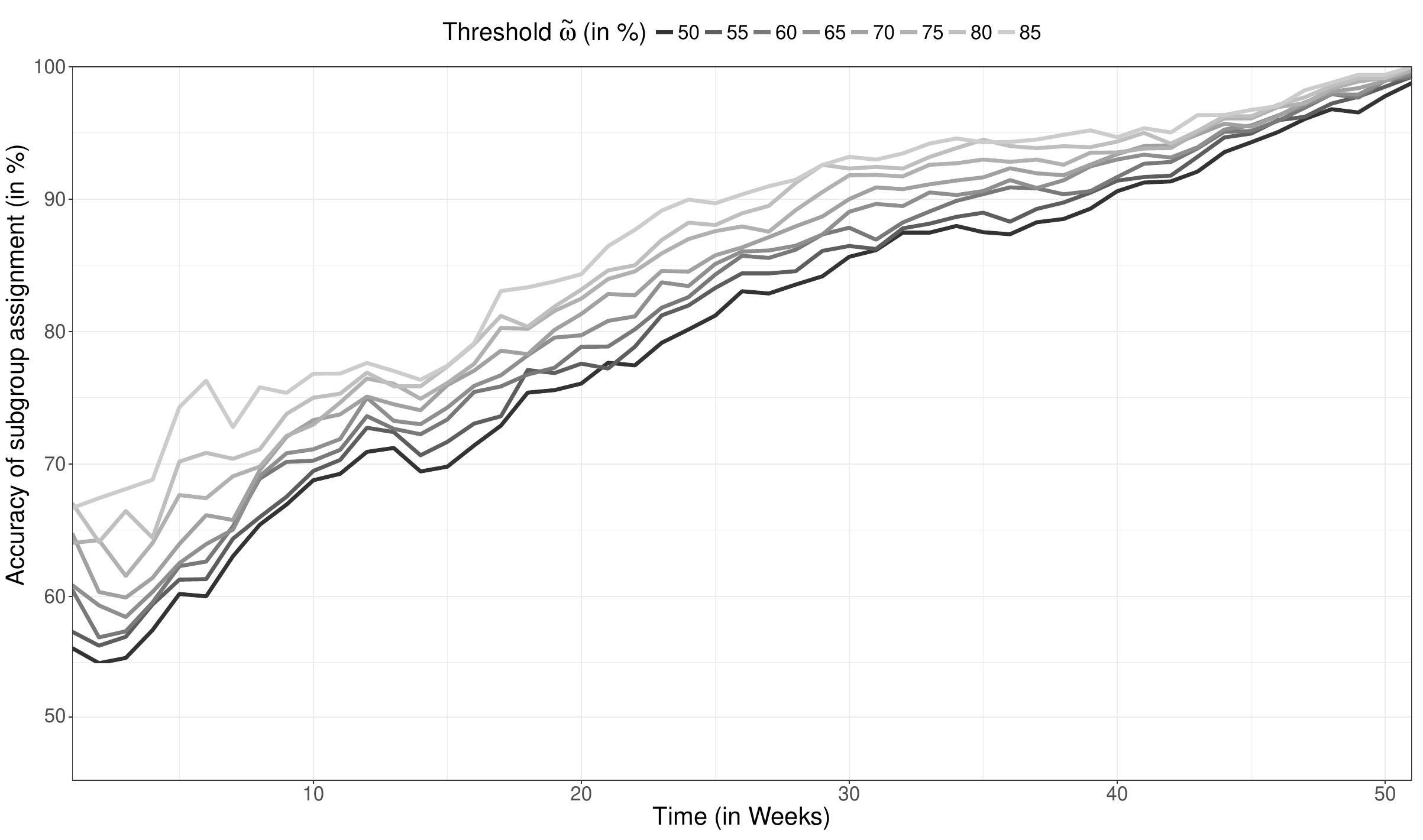}
	\caption{Accuracy of online subgroup assignments over time. The plot shows the level of agreement between the matching at time step $t$ and the final subgroup assignment. We see that, across all thresholds, the accuracy increases, which shows the informational value of having access to longer trajectory data.} 
	\label{fig:accuracy_over_time}
\end{figure}

Fig.~\ref{fig:accuracy_over_time} compares the accuracy of subgroup assignments as a function of the observed trajectory length. This should shed light on the informational value of trajectory data (and thus the longitudinal monitoring). We thus measure the overlap between the assignment $G_{it}$ at time step $t$ and the final outcome $G_{iT}$ at the end of the study horizon. To account for the probabilistic nature of our subgroup assignment (\ie, the membership is given by a probability $\omega_i^k$), we draw upon different thresholds $\tilde{\omega}$ at which the accuracy is evaluated. Specifically, we only consider patients for whom the maximum subgroup probability $\max \omega_i^k$ exceeds the threshold $\tilde{\omega}$. Evidently, the accuracy increases over time, demonstrating the high informational value encoded in trajectory data. For instance, after week \num{20}, the accuracy for the \SI{65}{\percent}-threshold exceeds \SI{80}{\percent}. In general, the initial trajectories tend to entail a higher information density, while the additional gain with each week flattens out later. We further notice that the difference in accuracy between different thresholds remains relatively constant at around \num{10} percentage points until week \num{23}. Afterward, the difference shrinks rapidly, falling below \num{5} percentage points. This shows that the subgroup probabilities have converged and are no longer subject to fluctuations. 

\textbf{Value of longitudinal information:} The clear benefit of our \model in clinical practice is the ability to identify subgroups that are informative of the future trajectory pattern as it unfolds. Our work further demonstrates the operational value of longitudinal monitoring (\eg, through smart devices or smartphone apps). In particular, information on health trajectories allows to better capture the underlying disease dynamics and thus tailor treatment plans accordingly \citep{Allam.2021}. Longitudinal monitoring is especially relevant for chronic conditions or conditions likely to become chronic, as these burden patients over a long time period. Here, longitudinal information can yield new insights into the underlying disease dynamics and thus enable better strategies for providing effective care.

\clearpage

\section{Extended discussion}

\textbf{Limitations:} As with other medical research, our work has strengths but also limitations with opportunities for future research. (1)~Our study involves a large cohort of patients with standardized examinations and different patterns of progression. This heterogeneity is a result of the recruitment of patients with and without leg pain, only excluding those with pregnancy or serious pathology. However, we cannot extend our findings to the excluded patient cohorts. (2)~We draw upon an extensive set of \num{17} potential risk factors that were chosen together with our clinical experts. If desired, one can also estimate our model with other variables. Future research may also seek for novel, digital markers that can help to facilitate the subgroup assignment for incoming patients. (3)~We are aware that clinical research related to low back pain distinguishes between the terms  \textquote{prognostic factors} and \textquote{risk factors}. However, for consistency with earlier works in health modeling, we prefer the latter. (4)~Our framework focuses on the identification of clinically-relevant subgroups. Designing treatment plans for subgroups is beyond our work and covered by related research \citep[\eg,][]{Bertsimas.2017,Helm.2015,Feuerriegel.2024}. (5)~Future research may add by validating our subgrouping in other samples as well as by using the subgrouping for stratified treatment plans and demonstrating their effectiveness compared to non-stratified plans in randomized controlled trials.

\textbf{Practical considerations:} Some risk factors may be time-varying, and, if desired, such time-varying risk factors can simply be entered additionally as emissions, so that information from both $t=0$ and $t=1, \ldots, T$ can be leveraged. Furthermore, we expect that our model may need to be updated with new data but this happens after longer time intervals (e.g., after one or more years). Here, the estimation can be accelerated by initializing the MCMC sampling with the parameters from the previous fit, which will greatly improve the convergence of the Markov chains.

\end{document}